\documentclass{article}
\usepackage{subcaption}

\usepackage{microtype}
\usepackage{graphicx}
\usepackage{multirow}
\usepackage{booktabs} 
\usepackage{soul}

\usepackage{hyperref}
\usepackage{adjustbox}
\usepackage{tikz}
\usepackage[most]{tcolorbox}
\global
\usepackage{nicefrac}

\usepackage{alltt}
\tcbset{
  aibox/.style={
    width=474.18663pt,
    top=10pt,
    colback=white,
    colframe=black,
    colbacktitle=black,
    enhanced,
    center,
    attach boxed title to top left={yshift=-0.1in,xshift=0.15in},
    boxed title style={boxrule=0pt,colframe=white,},
  }
}
\newtcolorbox{AIbox}[2][]{aibox,title=#2,#1}

\usepackage[hang,flushmargin]{footmisc}

\usepackage[accepted]{contents/sty_files/icml2024}

\usepackage{amsmath}
\usepackage{amssymb}
\usepackage{mathtools}
\usepackage{amsthm}
\usepackage{multirow}

\theoremstyle{plain}
\newtheorem{theorem}{Theorem}[section]

\theoremstyle{definition}
\newtheorem{definition}[theorem]{Definition}

\theoremstyle{remark}

\newcommand\blfootnote[1]{%
  \begingroup
  \renewcommand\thefootnote{}\footnote{#1}%
  \addtocounter{footnote}{-1}%
  \endgroup
}

\usepackage[capitalize,noabbrev]{cleveref}

\usepackage{amsmath,amsfonts,bm}

\renewcommand{\epsilon}{\varepsilon}

\def\eqref#1{equation~\ref{#1}}

\def\ceil#1{\lceil #1 \rceil}

\def\1{\bm{1}}

\def\vc{{\bm{c}}}

\def\ve{{\bm{e}}}

\def\vh{{\bm{h}}}

\def\vp{{\bm{p}}}

\def\mI{{\bm{I}}}

\DeclareMathAlphabet{\mathsfit}{\encodingdefault}{\sfdefault}{m}{sl}
\SetMathAlphabet{\mathsfit}{bold}{\encodingdefault}{\sfdefault}{bx}{n}

\def\sD{{\mathbb{D}}}

\def\sG{{\mathbb{G}}}

\DeclareMathOperator*{\argmin}{arg\,min}

\icmltitlerunning{Privacy-Preserving Instructions for Aligning Large Language Models}

\makeatletter
\def\@fnsymbol#1{\ensuremath{\ifcase#1\or \dagger \or  \ddagger\or
   \mathsection\or  \text{*}\or \mathparagraph \or  \| \or **\or \dagger\dagger
   \or \ddagger\ddagger \else\@ctrerr\fi}}
\makeatother

\renewcommand{\thefootnote}{\fnsymbol{footnote}} 

\begin{document}

\twocolumn[
\icmltitle{Privacy-Preserving Instructions for Aligning Large Language Models}

\begin{icmlauthorlist}
\icmlauthor{Da Yu$^{\dagger}$ }{}
\icmlauthor{Peter Kairouz$^{\ddagger}$ }{}
\icmlauthor{Sewoong Oh$^{\ddagger}$ }{}
\icmlauthor{Zheng Xu$^{\ddagger}$ }{}
\end{icmlauthorlist}

\vskip 0.4in
]

\begin{abstract}
Service providers of large language model (LLM) applications collect user instructions in the wild and use them in further aligning LLMs with users' intentions. These instructions, which potentially contain sensitive information, are annotated by human workers in the process. This poses a new privacy risk not addressed by the typical private optimization.  
To this end, we propose using synthetic instructions to replace real instructions in data annotation and model fine-tuning. 
Formal differential privacy is guaranteed by generating those synthetic instructions using privately fine-tuned generators. 
Crucial in achieving the desired utility is our novel filtering algorithm that matches the distribution of the synthetic instructions to that of the real ones. 
In both supervised fine-tuning and reinforcement learning from human feedback, our extensive experiments demonstrate the high utility of the final set of synthetic instructions by showing comparable results to real instructions. In supervised fine-tuning, models trained with private synthetic instructions  outperform leading open-source models such as Vicuna.
\footnotetext[1]{Sun Yat-sen University, the work was done when Da Yu was an intern at Google Research. \texttt{yuda3@mail2.sysu.edu.cn}. }
\footnotetext[2]{Google Research, alphabetical order.\\ \texttt{\{kairouz,sewoongo,xuzheng\}@google.com}.\\}
\blfootnote{\textit{Proceedings of the $\mathit{41}^{st}$ International Conference on Machine Learning}, Vienna, Austria. PMLR 235, 2024. Copyright 2024 by the author(s).}
\end{abstract}

\renewcommand{\thefootnote}{\arabic{footnote}}  

\section{Introduction}
\label{sec:intro}

\begin{figure}
\centering
\includegraphics[width=0.45\textwidth]{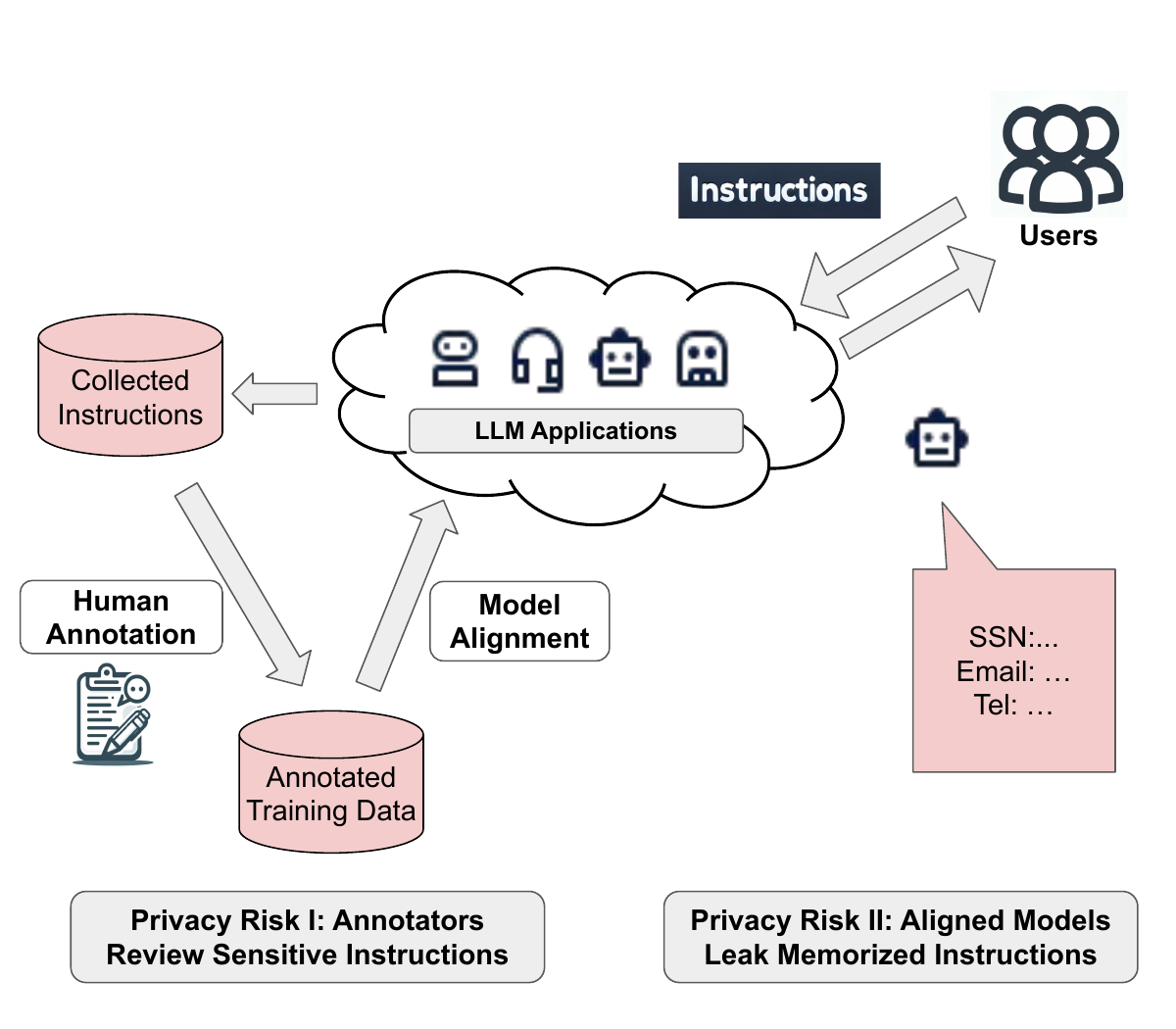}
\caption{Privacy concerns of collecting and training with user instructions in LLM applications.
}
\label{fig:two_threats}
\end{figure}

Training models to follow instructions is fundamental to the development of large language model (LLM) applications \citep{wei2021finetuned,chung2022scaling,ouyang2022training}. Instructions in the form of prompts provided by users during interactions with LLM applications are used to refine the backbone models \citep{OpenAIPrivacyPolicy}. These instructions are often reviewed and annotated by human annotators to create the training data for this refinement process \citep{washingtonpost_labeler}.
However, user-provided instructions can be highly sensitive since users may disclose personal information to elicit accurate responses. To illustrate this, we conducted a quantitative analysis on approximately twelve thousand real-world user instructions collected from the Chatbot Arena ~\citep{ChatbotArenaConversations}. This analysis begins with an LLM-based detection, followed by human verification, as detailed in Appendix~\ref{sec:instructions_are_sensitive}. Our analysis finds at least thousand of sensitive text pieces. A few examples of sensitive instructions are shown in Figure~\ref{fig:true_positives}. 

Given the sensitive nature of user instructions, we identify the following privacy risks in the pipeline of aligning LLMs with instructions, as illustrated in Figure~\ref{fig:two_threats}. During the data collection stage, potentially sensitive instructions are collected and presented to human annotators. In the subsequent training and deployment stages, trained models may memorize training data, including sensitive instructions, and inadvertently leak this memorized data during deployment. While the privacy risk of model memorization has been discussed in previous studies~\citep{carlini2021extracting,carlini2023extracting,nasr2023scalable,wu2024privately}, to the best of our knowledge, this work is the first to discuss the privacy risks associated with the annotation process.

One common approach to mitigate privacy concerns is to detect sensitive information in user instructions by either human annotation or automatic tools, and then remove this information from the model training process. For example, \citet{ouyang2022training} ask annotators to remove personally identifiable information (PII) from instructions before using them for model training \citep[Footnote A]{InstructGPTBlog}. However, there are several concerns that cannot be fully addressed by PII removal alone.  First, detecting PII by human annotators exposes sensitive instructions to annotators. Second, PII detection by automatic tools has high false negative rates. \citet{bubeck2023sparks} show that GPT-4, despite far surpassing existing automatic tools for PII detection, cannot detect all pieces of personal information in over 20\% of the documents in the Text Anonymization Benchmark \citep{pilan2022text}. Third, content that appears safe and cannot be identified as PII during the detection process may still reveal a user's identity if the adversary has side information \citep{narayanan2008robust}. To address these limitations, we develop a framework for aligning LLMs with  instructions while providing differential privacy (DP) guarantees \citep{dwork2006calibrating}, which theoretically limit the influence of any individual training sample and offer strong protection against empirical attacks \citep{rahman2018membership,carlini2019secret,lowy2024does}.

The standard method for fine-tuning LLMs on private data is to replace standard optimizers with their DP variants \citep{yu2021large,li2022large,bu2022differentially,wu2024privately}. However, merely ensuring that the fine-tuned model satisfies DP can only address the risk of memorization; 
there is no protection in the data collection stage when the user instructions are exposed to annotators. To simultaneously address both privacy risks, human annotators and model memorization, we propose using synthetic instructions (generated from privately fine-tuned language models)
as substitutes for real user instructions in the alignment pipeline.

{\bf Contributions.}
Our main contributions are summarized as follows\footnote{Our code is available at \url{https://github.com/google-research/google-research/tree/master/dp_instructions}.}.

\begin{figure*}[htb]
\centering
\includegraphics[width=1.0\textwidth]{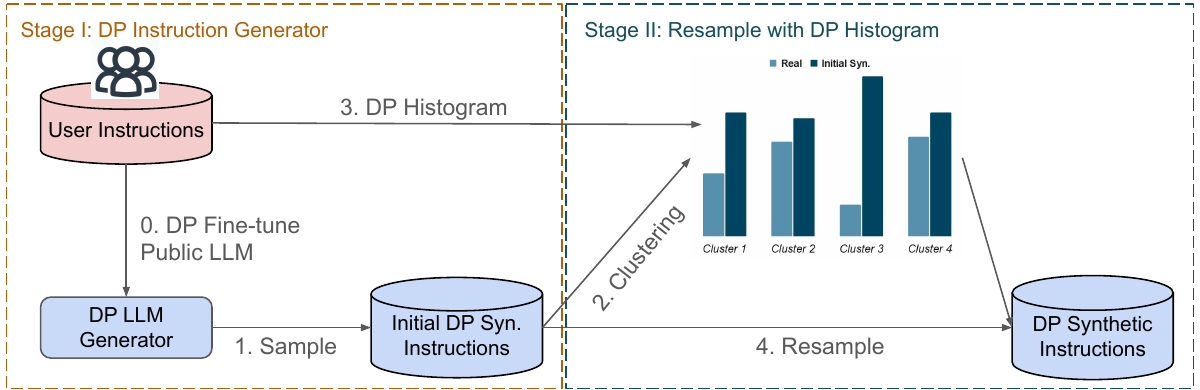}
\caption{Our two-stage framework for privately generating high-quality synthetic instructions.}
\label{fig:framework}
\end{figure*}

We introduce a two-stage framework for privately generating high quality synthetic instructions, which  can be seamlessly incorporated into the training pipeline of instruction-following models (\cref{fig:framework}). First, an LLM pre-trained on public data is privately fine-tuned on user-provided  instructions, from which a large set of synthetic instructions are sampled (\cref{alg:dp_syn}). We adopt state-of-the-art  strategies for DP fine-tuning including initializing with a large pre-trained model and parameter-efficient fine-tuning \citep{yu2022differentially,yue2022synthetic,kurakin2023harnessing}. 

There is still a large distribution shift between the user-provided instructions and the initial set of private machine-generated  instructions. To bridge this gap, we introduce the second stage of a novel private resampling algorithm (\cref{alg:dp_filt}); we select a subset of the initial private synthetic instructions in order to better match the real user instructions. Specifically, we first use a private histogram to approximate the distribution of the real instruction dataset, where each bin corresponds to each cluster learned from synthetic instructions, for example, representing coding questions for a programming language. The private histogram is then used to filter the initial synthetic instructions. The construction of the histogram is inspired by \citet{pillutla2022mauve} which uses similar cluster-and-histogram strategy to measure divergences between two datasets. 

We demonstrate the performance of the proposed approach on standard methods of aligning language models: for supervised instruction fine-tuning on LLaMA \citep{touvron2023llama1} 7B and 13B models and for reinforcement learning from human feedback on  1.3B Phi-1.5 model \citep{li2023textbooks}. We evaluate them on instructions from the LMSYS-Chat-1M dataset \citep{zheng2023lmsys} and the AlpacaEval benchmark \citep{dubois2023alpacafarm}. 
In supervised fine-tuning, training a 7B LLaMA model with resampled DP synthetic instructions leads to an 8.6\% relative improvement compared to using the unfiltered initial DP synthetic instructions. The improvement is measured in the win-rate against a baseline model. In reinforcement learning from human feedback, our private approach is comparable to using real instructions without any privacy guarantees.

{\bf Overview.}
We provide background on fine-tuning LLMs on instructions in \cref{sec:pre} and discuss privacy risks involved. 
A new framework for generating high-quality synthetic instructions with DP is presented in \cref{sec:algorithm}. 
Empirical results demonstrating the utility of those synthetic instructions are provided in \cref{sec:exps}.
We end with a conclusion in \cref{sec:conclusion}.

\section{Privacy Risks and Background}
\label{sec:pre}

\begin{figure*}[htb]
\begin{AIbox}[width=1.0\textwidth]{}
\parbox{1\textwidth}{\scriptsize\begin{alltt} 

\textbf{[Type: creative writing.]}  You are a grant writing super hero. Write up a basic grant for \hl{[Organization Name]} in \hl{[Location]}, to get 25,000 dollars for a new \hl{[Event Name]}. 


\textbf{[Type: information extraction.]}  \hl{[Medical Report]}

can you find the Date of servvice and provider name 


\textbf{[Type: brainstorming.]}  I am a \hl{[Job Title]} working in a \hl{[Country Name]} company that ...... At the same time, since \hl{[Date]}, I follow studies at \hl{[Program Name]}. ...... I'd like the AI to support specifically in defining key questions to ...... 


\textbf{[Type: document rewriting.]}  i need you to act as a professional human translator, native speaker in English Italian ...... the following is a business email, translate it in idiomatic italian with a formal tone: Hello xxx,
I am xxx, \hl{[Job Title]} of \hl{[Organization Name]}. I am contacting you to see if you would be interested in bringing your creative vision to \hl{[Location]}.
Based in \hl{[Location]}, we are ...... 

\end{alltt}}
\end{AIbox}
\caption{Samples of instructions containing personal information. We mask the sensitive texts  to protect user privacy.}
\label{fig:true_positives}
\end{figure*}

We present necessary background and closely related prior work. Additional related work is discussed in Appendix~\ref{apdx:more_related_work}.

\textbf{Privacy risks in training LLMs to follow instructions.}  Aligning LLMs with user instructions is crucial for ensuring accurate and contextually relevant responses, which in turn significantly enhances the user experience \citep{ziegler2019fine, ouyang2022training}. When  aligning pre-trained LLMs, a combination of the following two popular methods is used: $(i)$  supervised fine-tuning, which utilizes (instruction, answer) pairs to minimize the  cross-entropy loss, and $(ii)$ reinforcement learning from human feedback, where a reward model is trained on human preferences in the form of (instruction, \{answer\}, preference) to evaluate and score answers. Real-world user instructions are necessary in both alignment techniques, which might contain sensitive personal information. However, these instructions are sent to human annotators to get the answers in supervised fine-tuning and to get preferences in reward modeling \citep{ouyang2022training,touvron2023llama2}.  This introduces privacy concerns referred to as  Privacy Risk I in \cref{fig:two_threats}. The standard practice of keeping the alignment dataset proprietary is not sufficient to preserve the privacy of the participants. 

Only recently has it become possible to identify this vulnerability, following the release of several datasets featuring real-world user interactions with instruction-following models \citep{zheng2023lmsys,zhao2024inthewildchat}. We focus on two such datasets collected from the Chatbot Arena  \citep{zheng2023lmsys,ChatbotArenaConversations}, carefully adhering to their terms of use, and study privacy implications of using real-world instructions.

For the Chatbot Arena Conversations dataset \citep{ChatbotArenaConversations}, the authors mention that ``To ensure the safe release of data, we have made our best efforts to remove all conversations that contain personally identifiable information (PII)." Despite that, we identified numerous sensitive pieces of information in the publicly available dataset, which demonstrates how unreliable PII detection is and how vulnerable the instruction datasets are.  
The detailed PII analysis of the Chatbot Arena Conversations dataset is provided in \cref{sec:instructions_are_sensitive} and sample instructions containing personal information are shown in \cref{fig:true_positives}.

\textbf{Private deep learning.} We use  ($\varepsilon$,$\delta$)-DP \citep{dwork2006our} defined as,

\begin{definition}[$(\varepsilon,\delta)$-Differential Privacy]
  A randomized algorithm $\mathcal{M}$ satisfies  ($\varepsilon$,$\delta$)-DP for $\sD$ if for any two neighboring datasets $\sD$, $\sD'$ and for all $\mathcal{S}\subset \text{Range}(\mathcal{M})$: 
\[
\textstyle{\Pr[\mathcal{M}(\sD) \in \mathcal{S}] \leq e^{\epsilon}\Pr[\mathcal{M}(\sD') \in \mathcal{S}] +\delta}
.\]
\end{definition}
Neighboring datasets are defined by adding/removing one sample, where each sample constitutes a single user instruction (text prompt), i.e., example-level DP \citep[Section 5.1]{ponomareva2023dp}. $\mathcal{M}$ is a (randomized) training algorithm that outputs trained model weights.  The most common approach for private deep learning is to  clip and add noise to each gradient update \citep{song2013stochastic,abadi2016deep,kairouz2021practical}. We use a variant of DP-SGD \citep{abadi2016deep} with Adam optimizer, which is called DP-Adam \citep{li2022large}.

\textbf{Differentially private synthetic data.} A widely adopted method for generating  private synthetic text in practice is to train a language model on private data with DP,  and then repeatedly sample the DP model to  generate synthetic text sequences~\citep{augenstein2019generative,bommasani2019towards,putta2022differentially,mattern2022differentially,yue2022synthetic, kurakin2023harnessing}. In this work, we follow this general approach to generate initial synthetic instructions. Importantly, we extend beyond the scope of prior studies, which primarily focus on simple tasks such as text classification, by investigating the use of private synthetic text in the training of instruction-following models. Moreover, we introduce a novel filtering algorithm  that substantially reduces the distributional gap between real and private synthetic texts.

\textbf{Privacy-preserving In-Context Learning (ICL).} In addition to aligning LLMs through fine-tuning, ICL is another paradigm for adapting LLMs to specific domains. ICL improves a model's performance in a target domain by providing instructions and few-shot demonstrations directly in the prompt \citep{brown2020language}. To protect the privacy of in-context prompts \citep{duan2023privacy}, a recent line of research focuses on generating differentially private prompts for in-context learning \citep{duan2024flocks,wu2024privacy,tang2024privacy,hong2024dp}. Our work differs from DP-ICL in two key aspects. Firstly, regarding the threat model, DP-ICL protects sensitive information within in-context prompts, while our approach protects the privacy of training data against both model memorization and exposure to human annotation. Secondly, in terms of algorithms, our framework produces a large set of synthetic instructions rather than few-shot demonstrations, catering to the requirement of a larger set of data in fine-tuning for LLM alignment.

\section{Generating Synthetic Instructions with Differential Privacy}
\label{sec:algorithm}

To mitigate the privacy concerns in aligning LLMs, we propose a novel privacy-preserving approach by generating synthetic instructions with DP (see Figure~\ref{fig:framework} for an overview). First, an LLM generator is  fine-tuned with DP on user-provided instructions, which is subsequently used to generate a large set of initial {\em synthetic} instructions (Algorithm~\ref{alg:dp_syn}). Next, these initial synthetic instructions are resampled according to a private histogram to approximately match the distribution of the real instructions (Algorithm~\ref{alg:dp_filt}).
We carefully allocate the privacy budget in Algorithms~\ref{alg:dp_syn} and~\ref{alg:dp_filt}, and the formal DP guarantees of the entire process can be composed \citep{kairouz2015composition,steinke2022composition}. The generated DP synthetic  data can be used for aligning LLMs and the same DP guarantees are applied because of the post-processing property of DP \citep{algofound}.

\subsection{Stage 1: DP Instruction Generator}
\label{subsec:pre_syn}

We take an LLM, pre-trained on public data, and fine-tune on private instruction data with DP. This is used to generate samples that comprise a large set of initial instruction dataset, see Algorithm~\ref{alg:dp_syn}.   Concretely, LLaMA models \citep{touvron2023llama1} are privately fine-tuned on user instructions with LoRA \citep{hu2022lora}. LoRA has been shown to have superior privacy-utility trade-offs in fine-tuning of DP text generators \citep{kurakin2023harnessing}. Previous research has shown that DP training is effective at preventing memorization of training data \citep{guo2022bounding, stock2022defending, yue2022synthetic}. In Appendix~\ref{sec:canary}, we provide further evidence by by showing that text generative models trained without DP can memorize training samples and leak them during inference, whereas DP-trained models do not.

\begin{algorithm}[htb]
   \caption{Train DP Generator to Synthesize Instructions.}
   \label{alg:dp_syn}
\begin{algorithmic}[1]
  \STATE {\bfseries Input:} real instructions $\sD_{real}$, pre-trained model $\theta$, privacy parameters $(\varepsilon_{1},\delta_{1})$, number of generations $M$.

  \medskip

  \STATE Fine-tune $\theta$ on $\sD$ with DP-Adam and $(\varepsilon_{1}, \delta_{1})$-DP.
  
  \STATE Sample from the fine-tuned model for $M$ times to generate $M$ synthetic instructions ($\sD_{syn}$).

  \RETURN $\sD_{syn}$

\end{algorithmic}
\end{algorithm}

Since the fine-tuned model itself is private, we can sample as many instructions as we want. 
However, there is a non-negligible distributional gap between the initial synthetic instructions and real instructions, which increases as the DP guarantee strengthens, likely due to the noise added in DP training. The results in Table~\ref{tbl:mauve_main} in Section~\ref{subsec:mauve} illustrate this phenomenon. We then present a DP resampling algorithm to improve the quality of initial synthetic instructions.

\subsection{Stage 2: Resample with DP Histogram}
\label{subsec:filter_syn}

Let $\sD_{real}$ be a set of real instructions of size $N$, $\sD_{syn}$ be a set of initial synthetic instructions of size $M$ ($M>N$), and $\phi$ be an embedding model that maps an instruction into a real-valued vector. Our goal is to select $\sD_{syn}^{'}\subseteq \sD_{syn}$ whose distribution more closely aligns with that of $\sD_{real}$ in the embedding space.

The pseudocode of our resampling process is presented in Algorithm~\ref{alg:dp_filt}. We first  groups synthetic instructions, $\sD_{syn}$, into $K$ clusters, $\{\vc_{j}\}_{j=1}^K$, using k-means. Clustering synthetic instructions does not incur additional privacy cost because it constitutes a post-processing step of Algorithm~\ref{alg:dp_syn}. The cluster centroids of $\sD_{syn}$ are used to  create a histogram of real instruction samples in $\sD_{real}$ where each bin corresponds to each cluster.  This histogram, $\vh$, is made private by adding appropriate Gaussian noise to each bin. The synthetic data,  $\sD_{syn}$, is resampled to match this private histogram. \cref{tbl:mauve_main} in Section~\ref{subsec:mauve} demonstrates that the filtered  $\sD_{syn}'$ is closer to $\sD_{real}$ than the initial $\sD_{syn}$, which in turn improves the performance of the model fine-tuned on the filtered dataset. Further studying the clusters of $\sD_{syn}$ reveals that, for example, the top voted cluster contains instructions of coding questions in Python as shown in Figure~\ref{fig:cluster_samples} of Appendix~\ref{apdx:additional_plots}. On the other hand, clusters receive least votes do not exhibit any coherent semantics.

Algorithm~\ref{alg:dp_filt} reduces the gap between real and synthetic datasets by aligning their histograms. The idea of clustering samples in a dataset into a histogram is inspired by MAUVE \citep{pillutla2021mauve, pillutla2022mauve}. MAUVE is a metric designed to quantify the discrepancy between synthetic and real data. It first selects methods to represent the distributions of both synthetic and real datasets and then measures the divergence frontiers \citep{djolonga2020precision}. One specific method involves clustering the samples in a dataset into a histogram and using this histogram as the approximated distribution.

\begin{algorithm} [tb]
   \caption{Resample Synthetic Data with DP Histogram.}
   \label{alg:dp_filt}
\begin{algorithmic}[1]
  \STATE {\bfseries Input:} initial synthetic instructions $\sD_{syn}$, real instructions $\sD_{real}$, embedding model $\phi$, number of clusters $K$, privacy budget $(\varepsilon_{2},\delta_{2})$-DP, number of target synthetic samples $T$.

  \STATE Compute the embeddings of synthetic instructions $\{\ve^{syn}_{i}\}_{i=1}^{M}$ and real instructions $\{\ve^{real}_{i}\}_{i=1}^{N}$ with $\phi$.
  \STATE Run k-means to cluster  $\{\ve^{syn}_{i}\}$ into $K$ groups. Let the $i_{th}$ group be $\sG_{i}$ and the corresponding centroid be $\vc_{i}$.
  \STATE Initialize a $K$-dimensional histogram $\vh$.
  \STATE \textsl{//Each real instruction chooses the nearest centroid.}
  \FOR{$i=1$ {\bfseries to} $|\sD_{real}|$}
    \STATE $\gamma=\argmin_{j=1}^{K} |\ve_{i}^{real}-\vc_{j}|$.
    \STATE $\vh[\gamma] = \vh[\gamma]+1$.
  \ENDFOR
  \STATE \textsl{//Privatize the histogram and compute the densities.}
  \STATE $\tilde \vh=\vh+z, z\sim \mathcal{N}(0,\sigma^{2}\mI_{K\times K})$, where $\sigma$ is set to guarantee $(\varepsilon_{2},\delta_{2})$-DP.
  \STATE $\vp=\tilde \vh/|\sD_{real}|$.
  \STATE \textsl{//Resample initial synthetic instructions.}
  \FOR{$i=1$ {\bfseries to} $K$}
    \IF{$|\sG_{i}|<\ceil{T*\vp}$}
        \RETURN \textsl{`Need more initial samples.'}
    \ELSE
        \STATE Uniformly sample  $\max(\ceil{T*\vp},0)$ synthetic instructions from $\sG_{i}$. Let $\hat \sG_{i}$ be the sampled subset.
    \ENDIF
  \ENDFOR 
  \RETURN $\cup_{i=1;K} \hat \sG_{i}$.
\end{algorithmic}
\end{algorithm}

\begin{figure}[htb]
\centering
\includegraphics[width=0.49\textwidth]{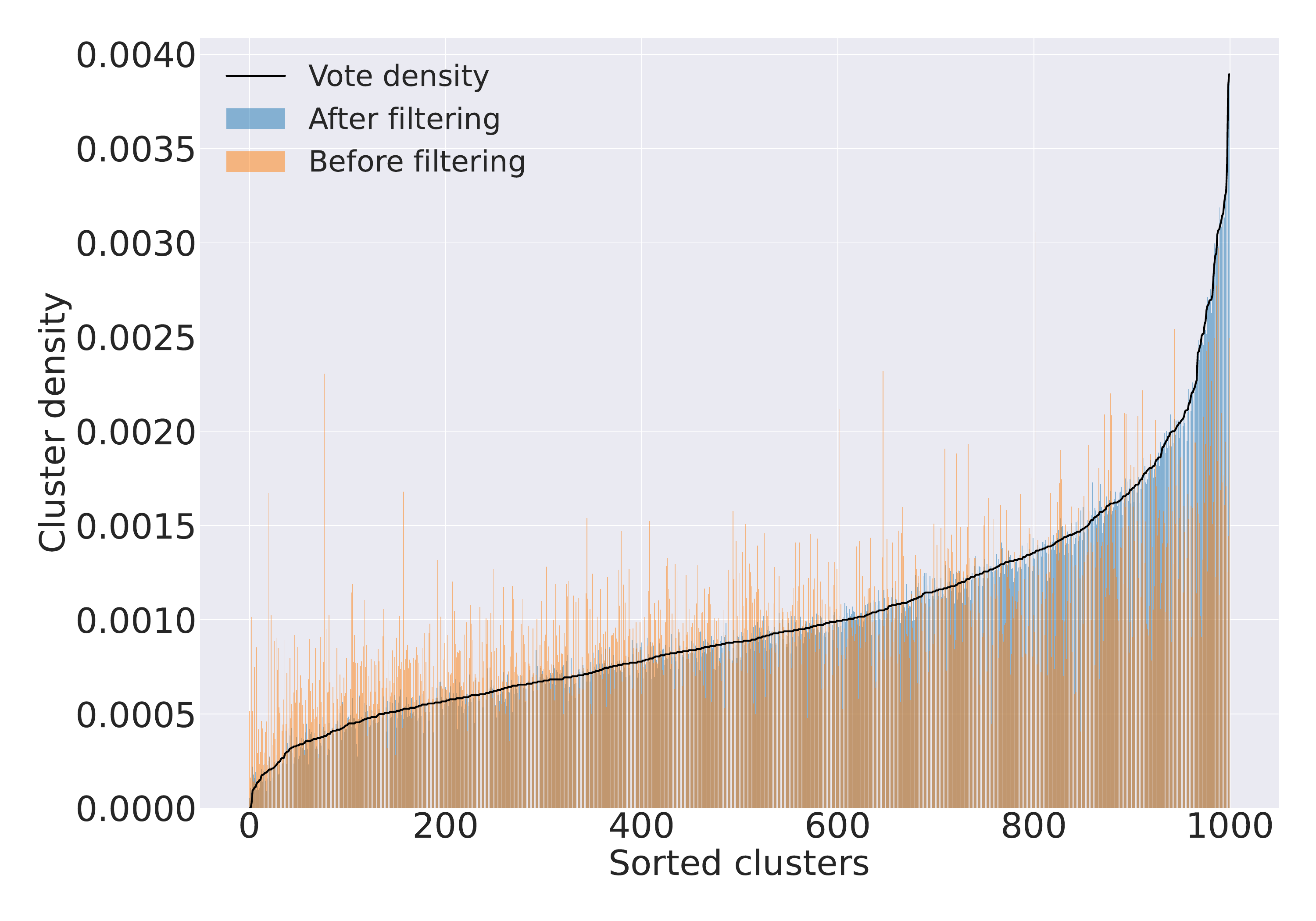}
\caption{Probability densities of clusters of synthetic instructions. The black line shows the sorted votes from real samples (before noising). The filtering process aligns the distribution of synthetic instructions with that of real instructions. }
\label{fig:cluster_densities}
\end{figure}

The size of the histogram, $K$, traverses  a trade-off between signal and noise: a large $K$ gives a fine-grained characterization of the real instruction distribution, but also  makes the histogram harder to privatize because of the increased dimension  and reduced votes in each cluster. 
As the votes from real instructions sum up to $N$, a large $K$ reduces the number of votes per-cluster. Therefore,  adding proper noise to the histogram for meaningful DP guarantees would result in an inaccurate estimation of the real instruction distribution due to poor signal-to-noise ratio. One good practice  for determining the value of $K$ is to ensure that the average number of votes per cluster is sufficiently robust to DP noise. In our experiments, we investigate the quality of resampled synthetic data, measured by MAUVE, for varying $K$. As depicted in  Figure~\ref{fig:vary_k_sigma} of Section~\ref{subsec:mauve},  while the resampling process enhances  the quality of synthetic data for all choices of $K$, the improvement initially increases with $K$ but beyond a certain point, it begins to either plateau or decline.

We illustrate the resample process in Figure~\ref{fig:cluster_densities} with clusters $K=1000$ and noise $\sigma=10$. We fine-tune LLaMA 13B \citep{touvron2023llama1} with DP-Adam and sample one million synthetic instructions from the fine-tuned model.  The overall privacy cost for fine-tuning and resampling is $(5.98,5\times 10^{-7})$-DP. Other implementation details, such as the choice of the embedding model, are in Section~\ref{sec:exps}. After filtering, there are around 310 thousand synthetic instructions left. Figure~\ref{fig:cluster_densities} shows that the empirical clustered distribution  of synthetic instructions after filtering matches that of real instructions. 

It is worth noting that Algorithm~\ref{alg:dp_filt} has a much lower privacy cost compared to Algorithm~\ref{alg:dp_syn}. It only releases a  histogram once which can tolerate much larger noise. 
For example, as depicted in Table~\ref{tbl:mauve_main} of Section~\ref{subsec:mauve}, selecting $\epsilon_1=5.94$ and $\epsilon_2$ small enough that the end-to-end privacy is only $\epsilon=5.98$ is already enough to see significant gain of filtering.

\section{Experiments}
\label{sec:exps}

We conduct extensive experiments to evaluate our algorithms. In Section~\ref{subsec:mauve}, we assess the distributional gap between synthetic and real instructions. We then investigate the utility of synthetic instructions in both supervised instruction fine-tuning (Section~\ref{subsec:sft}) and reinforcement learning from human feedback (Section~\ref{subsec:rlhf}).

\subsection{Setup for Generating Synthetic Instructions}

\textbf{Instruction datasets.} We use the LMSYS-Chat-1M dataset  as the private dataset \citep{zheng2023lmsys}, which has one million conversations between real users and  instruction-following models collected from April to August 2023. We only use the user instructions and disregard the corresponding machine-generated responses. This data is first preprocessed  by deduplication and filtering out non-English instructions, among other steps, as detailed in Appendix~\ref{apdx:data_preprocess}. After preprocessing, approximately  200,000  instructions remain, which we refer to as Chatbot Arena instructions. The Chatbot Arena instructions are then divided into three subsets: 180,000 for the training set, 5,000 for the validation set, and the rest for the test set.

We compare the models fine-tuned on private instructions (with DP) with models fine-tuned on out-of-distribution public instructions (without DP). This comparison aims to demonstrate that, despite the abundance of public instruction datasets \citep{longpre2023flan, OpenOrca}, in-domain but sensitive user instructions are still essential for aligning LLMs.
For this, we take instructions from the FLAN dataset \citep{chung2022scaling,longpre2023flan}. FLAN comprises question/answer pairs derived from a  rich collection of classic NLP tasks. For our purposes, we retain only the instructions. We randomly sample 180,000  instructions from FLAN unless otherwise specified.

\textbf{Privacy accounting.} We consider two overall privacy budgets $\varepsilon \simeq 3$ and $\varepsilon \simeq 6$. The privacy parameter $\delta$ is set as $5\times 10^{-7}$, smaller than $ {0.1}/N$, where $N$=180,000 is the number of training samples.  We use the Privacy Random Variable accountant \citep{koskela2020computing,gopi2021numerical,ghazi2022faster} to compose the privacy costs of Algorithm~\ref{alg:dp_syn} and~\ref{alg:dp_filt}. The code snippet for privacy accounting is in Appendix~\ref{apdx:code_snippet_privacy_accounting}.

\textbf{Setup for Algorithm~\ref{alg:dp_syn}.} The DP generators are fine-tuned from LLaMA models \citep{touvron2023llama1}.  This ensures that no private data is leaked in the pre-training, since the cutoff date for the pre-training data of LLaMA is prior to the start of the collection of Chatbot Arena instructions. We tune the hyperparameters based on  MAUVE scores. Details regarding the hyperparameters for both fine-tuning and sampling are in Appendix~\ref{apdx:hyperparameters}.

 We undertake several  measures to generate high-quality initial synthetic instructions. Using MAUVE score between Chatbot Arena instructions and the initial synthetic instructions, we study the effect of model size with LLaMA 7B and 13B models. The results suggest that a larger model synthesizes better quality data (Appendix~\ref{apdx:llama7b_vs_13b}). 
One unexpected by-product of gradient clipping in DP fine-tuning is that  the clipping threshold has a significant impact on the length of synthetic instructions and, consequently, affects the MAUVE scores. The results of using varying clipping thresholds are in Appendix~\ref{apdx:vary_clipping}.  In the rest of this paper, generator models are fine-tuned from LLaMA 13B unless otherwise specified. We sample one million initial synthetic instructions from each fine-tuned model.

\textbf{Setup for Algorithm~\ref{alg:dp_filt}.} We cluster instruction datasets  using $K$-means clustering on the  Sentence-T5-base \citep{ni2021sentence} embeddings, 
using the Faiss package for an efficient implementation \citep{johnson2019billion}.  Unless otherwise specified, the  number of clustering bins $K$  and noise standard deviation $\sigma$ in Algorithm~\ref{alg:dp_filt} are set as $1000$ and $10$, respectively. Selecting a subset of 180,000 synthetic instructions requires around 743,000 and 574,000  initial synthetic samples for $\varepsilon\simeq 3$ and $\varepsilon\simeq 6$, respectively. A larger number of initial samples are required for higher privacy, i.e.,  $\varepsilon\simeq 3$, since sample quality is lower. 
\subsection{Measuring the Distributional Gap}
\label{subsec:mauve}

We compute the MAUVE score \citep{pillutla2021mauve,pillutla2022mauve} between synthetic and real instructions. MAUVE compares two text distributions using divergence frontiers \citep{kynkaanniemi2019improved, djolonga2020precision} and has been widely used to analyze the gap between synthetic text and real text \citep{yue2022synthetic, kurakin2023harnessing}. We use two representations of text distributions to compute the MAUVE scores: the unigram distribution, and the histogram of clustered text embeddings \citep{pillutla2021mauve}. Our setup for computing MAUVE scores is in Appendix~\ref{apdx:compute_mauve}.

\begin{table}[htb]

\renewcommand{\arraystretch}{1.25}
\centering
    \caption{MAUVE scores with Chatbot Arena instructions as the target dataset. Best scores of DP models are marked in bold.  Filtering  synthetic instructions with Algorithm~\ref{alg:dp_filt} improves the scores of private synthetic instructions by large margins.} \label{tbl:mauve_main}
      \smallskip
    \begin{adjustbox}{max width=0.45\textwidth}
        \begin{tabular}{l|c|c}
\hline

    & Unigram & Sentence-T5 \\
\hline
 FLAN (non-private, OOD)   &    $0.191$     &    $0.124$           \\
\hline \hline 
Synthetic ($\varepsilon=2.86$, no filtering) &    $0.932$     &  $0.893$ \\
\hline 
Synthetic ($\varepsilon=2.91$) &  $\textbf{0.958}$      & $\textbf{0.967}$  \\
\hline 
Synthetic ($\varepsilon=5.94$, no filtering) &     $0.942$    &  $0.912$ \\
\hline
Synthetic ($\varepsilon=5.98$) &   $\textbf{0.961}$      & $\textbf{0.975}$ \\
\hline \hline 
 Synthetic (non-private, no filtering)   &   $0.983$      &    $0.991$        \\
\hline
 Synthetic (non-private)   &   $0.989$      &    $0.999$        \\
\hline
        \end{tabular}
    \end{adjustbox}
\end{table}

\begin{figure*}[htb]
\centering
\begin{subfigure}{.49\textwidth}
    \centering
    \includegraphics[width=1.0\linewidth]{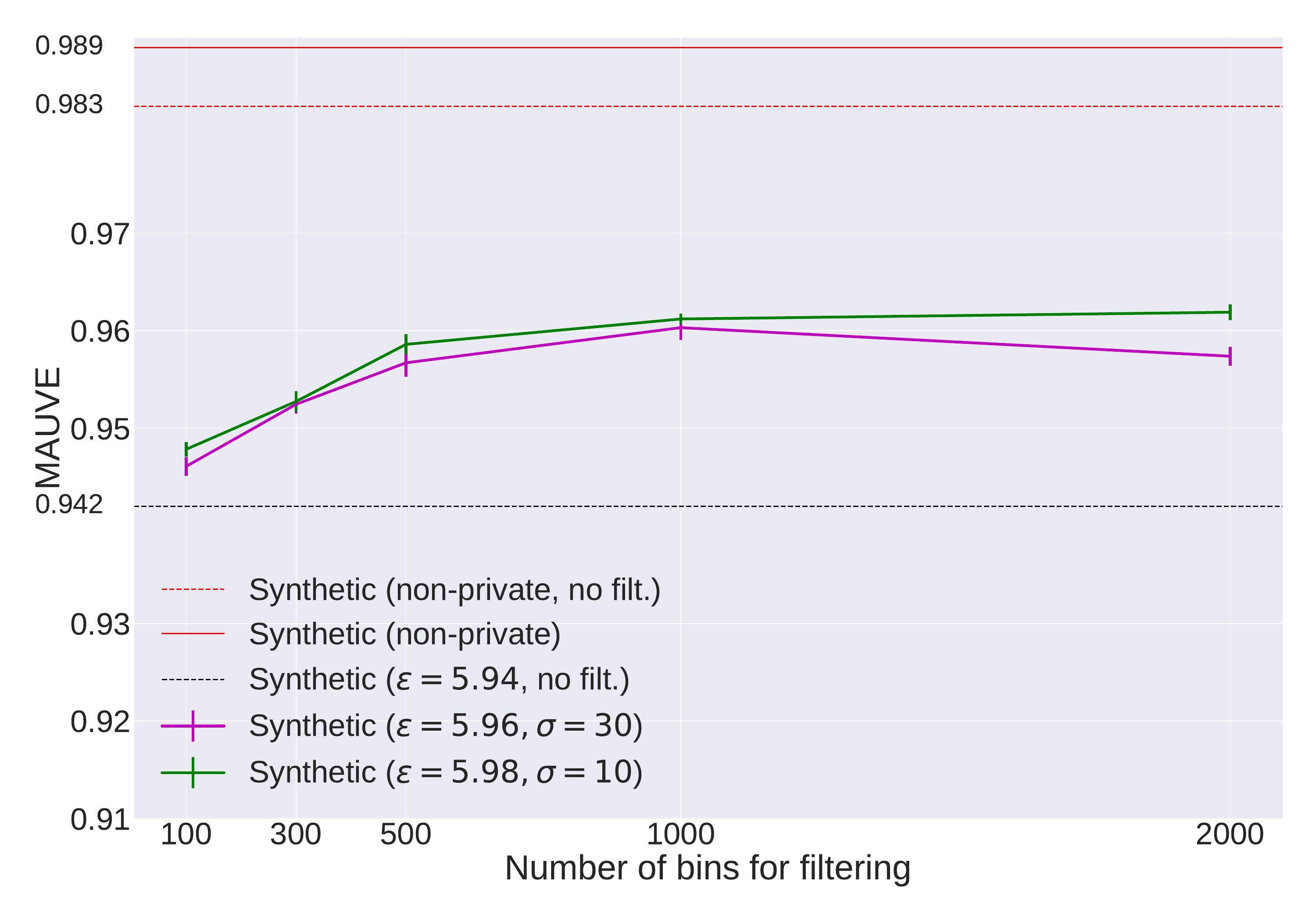}  
    \caption{MAUVE-Unigram}
\end{subfigure}
\begin{subfigure}{.49\textwidth}
    \centering
    \includegraphics[width=1.0\linewidth]{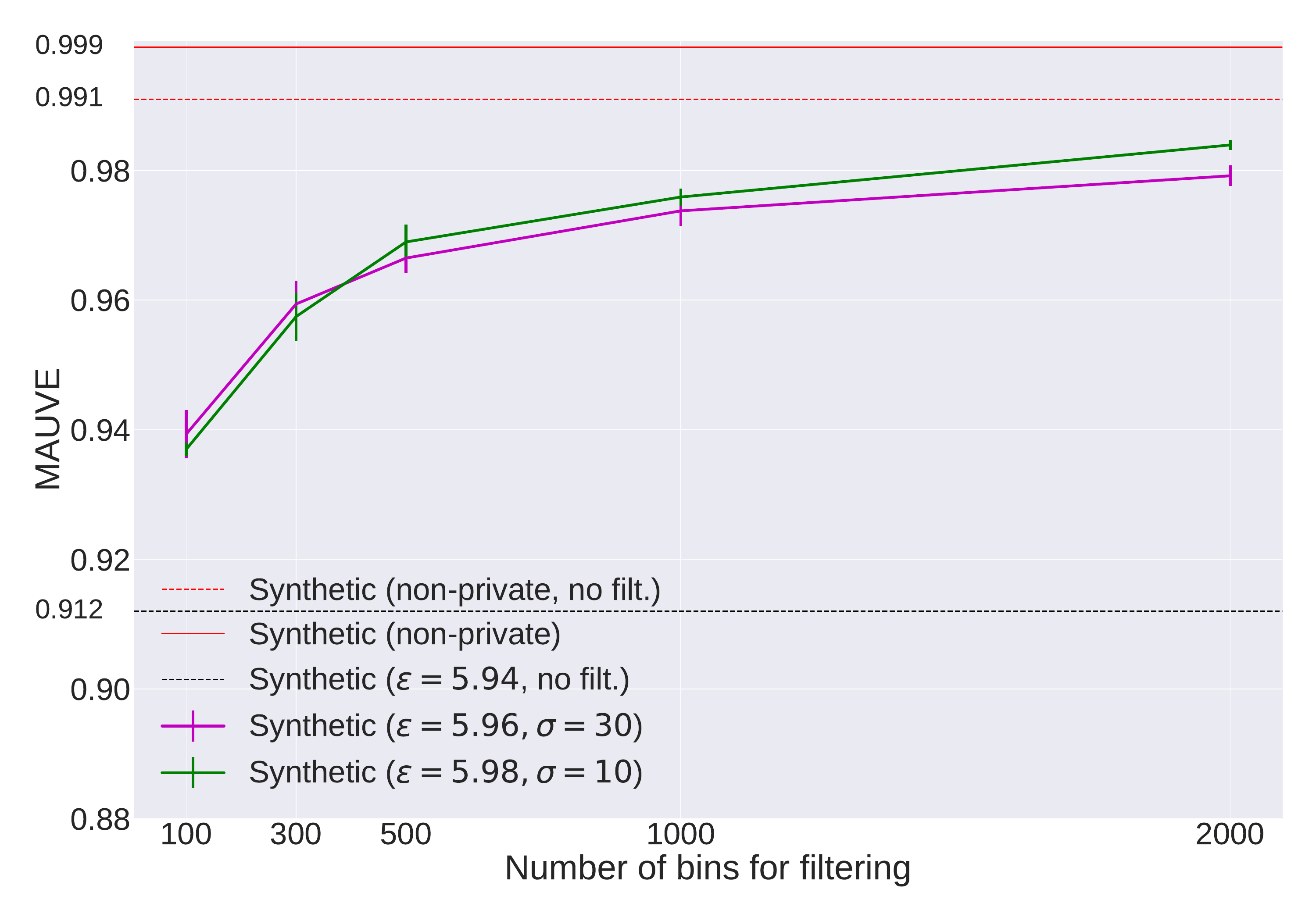}  
    \caption{MAUVE-Sentence-T5}
\end{subfigure}
\caption{Running Algorithm~\ref{alg:dp_filt} with different $K$ and $\sigma$. The MAUVE scores initially improve with an increase in $K$, then they either start to plateau or decline.}
\label{fig:vary_k_sigma}
\end{figure*}

Table~\ref{tbl:mauve_main} presents the main MAUVE scores.  We compute the scores with three random seeds and report the average. Filtering private  synthetic instructions with Algorithm~\ref{alg:dp_filt} improves MAUVE scores by large margins. We further demonstrate improvements in MAUVE scores for two other datasets, IMDB movie reviews and PubMed abstracts, to underscore the wide applicability of Algorithm~\ref{alg:dp_filt}.  Details on the results of the two additional datasets are in Appendix~\ref{apdx:other_datasets}. Additionally,  there is a significant distributional gap between FLAN instructions and real-world user instructions. This indicates that in-domain user instructions, while being sensitive, play a crucial role in LLM alignment and should be utilized in a privacy-preserving way.

\begin{figure}[htb]
\centering
\includegraphics[width=0.47\textwidth]{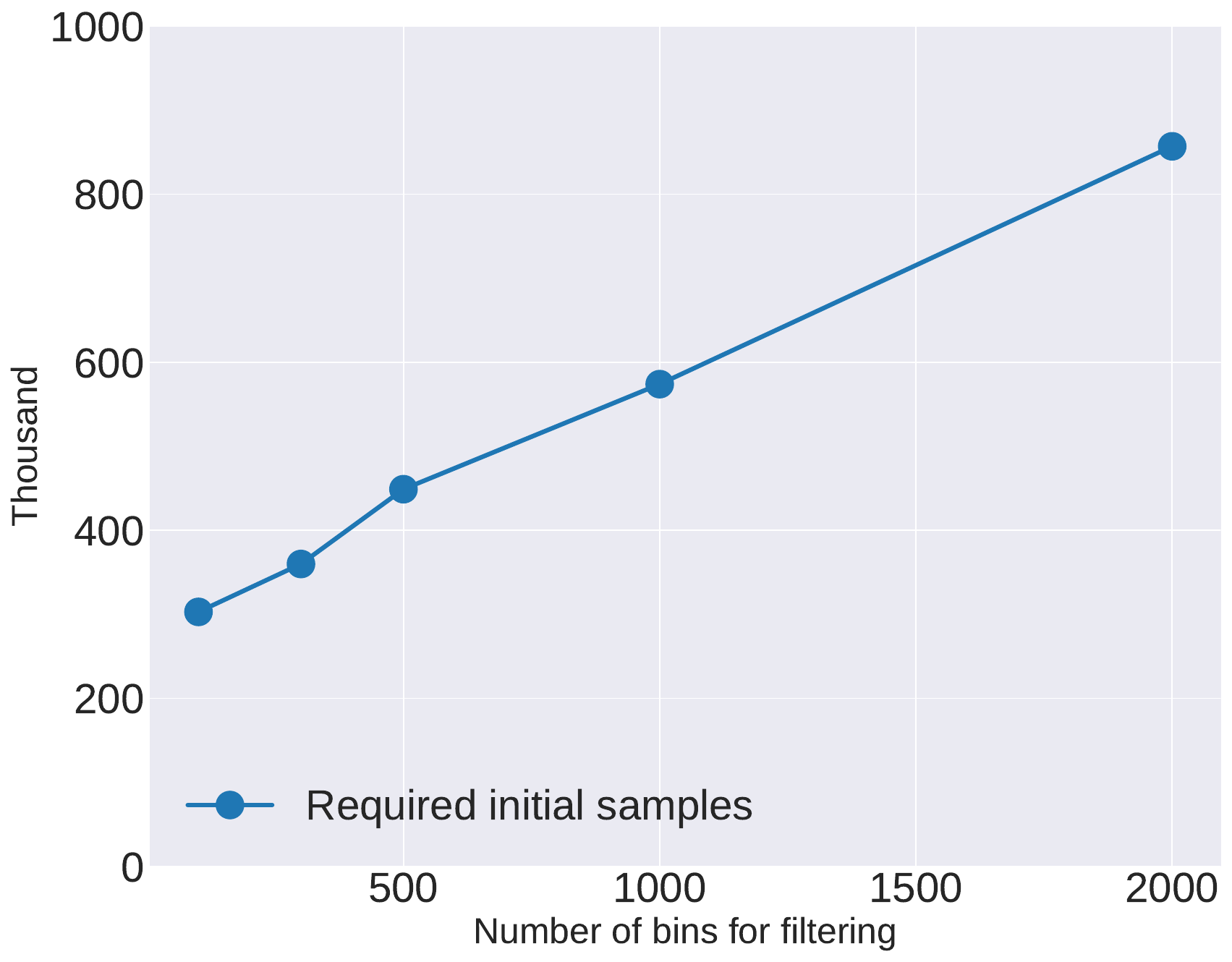}
\caption{The number of required initial samples for selecting a subset of 180 thousand increases with $K$. }
\label{fig:init_n_vary_k}
\end{figure}

We also run Algorithm~\ref{alg:dp_filt} with different values of $K$ and $\sigma$. The values of $K$ ranges from 100 to 2000, and the values of $\sigma$ are chosen from $\{10, 30\}$. The resulting MAUVE scores are presented in Figure~\ref{fig:vary_k_sigma}. The required  quantity of initial synthetic instructions as $K$ varies is presented in Figure~\ref{fig:init_n_vary_k}. The MAUVE scores initially improve with an increase in $K$. However, the scores begin to either plateau or decline for large values of $K$. This is likely because a larger value of $K$  weakens the aggregated counts, rendering the filtering process more sensitive to noise. Interestingly, the number of initial synthetic instructions required by Algorithm~\ref{alg:dp_filt} to select a subset of 180 thousand also increases with $K$. In the rest of this paper, we set $K=1000$ and $\sigma=10$ for filtering one million initial synthetic instructions. In Appendix~\ref{apdx:fewer_initial}, we run Algorithm~\ref{alg:dp_filt} with fewer initial synthetic instructions. The results suggest that filtering 500 thousand initial synthetic instructions results in inferior MAUVE scores compared to using one million initial instructions.  Notably, to select a subset of 180 thousand from 500 thousand initial samples, we have to use sampling with replacement  for large values of $K$.

\subsection{Supervised Fine-tuning}
\label{subsec:sft}

We now assess the utility of synthetic instructions by applying them in supervised instruction fine-tuning. One common way to generate the training data, consisting of  pairs of instructions and answers, is to ask human labelers to write the answers \citep{ouyang2022training,touvron2023llama2}. However, human annotation is notably costly both in time and finances. To circumvent this, we follow previous studies \citep{taori2023alpaca, vicuna2023} to employ GPT-3.5-Turbo for annotating answers for all instructions, including Chatbot Arena instructions, synthetic instructions, and FLAN instructions.  We then fine-tune LLaMA 7B or 13B models on the resulting datasets. All datasets contain 180 thousand pairs of instructions and answers unless otherwise specified.  Implementation details are in Appendix~\ref{apdx:implementation_details}.

We use GPT-4 as a judge to assess the instruction-following capabilities of the fine-tuned models. Specifically, we give GPT-4 one instruction accompanied by two responses generated by the two models being compared, and prompt it to choose a better response. The prompt we used is from \citet{alpaca_eval}. We then compute the average win-rate on a set of evaluation instructions. Recent research indicates that evaluating instruction-following models using GPT-4 is as effective as assessments conducted by humans \citep{dubois2023alpacafarm, zheng2023judging}. Furthermore, using GPT-4 for evaluation is scalable, enabling the efficient and cost-effective assessment of dozens of models.

\begin{table}[htb]
\small
\renewcommand{\arraystretch}{1.25}
\centering
    \caption{Win-rates on Chatbot Arena instructions over 3 seeds.  The left column shows the fine-tuning  sets (default size is 180K), except for Vicuna-v1.3 which is a public model and is not further fine-tuned.  Best win-rates of DP models are in bold. The win rate with filtering ($\varepsilon=2.91$) surpasses that without filtering ($\varepsilon=5.94$). } \label{tbl:winrate_arena}
      \smallskip
    \begin{adjustbox}{max width=0.45\textwidth}
        \begin{tabular}{l|c}
\hline
\multicolumn{2}{c}{\textbf{7B Models}} \\
\hline
FLAN (non-private)  & $50\%$  \\ 
\hline
Vicuna-v1.3 & $64.1\%$ ($\pm 0.61$)   \\ 
\hline 
Chatbot Arena (non-private) &   $68.9\%$ ($\pm 0.31$)  \\
\hline 
Chatbot Arena ($\varepsilon=5.94$) &  $60.7\%$ ($\pm 0.42$)  \\
\hline
Synthetic ($\varepsilon=5.94$, no filt.) &  $62.7\%$ ($\pm 0.34$)\\
\hline
Synthetic ($\varepsilon=2.91$) & $65.6\%$ ($\pm$0.36) \\
\hline
Synthetic ($\varepsilon=5.98$) & $67.8\%$ ($\pm 0.32$) \\
\hline
Synthetic  (300K, $\varepsilon=5.98$) & $\textbf{68.1\%}$ ($\pm 0.37$)  \\
\hline
\hline
\multicolumn{2}{c}{\textbf{13B Models}} \\
\hline
Vicuna-v1.3  & $72.8\%$ ($\pm 0.58$)   \\ 
\hline
Synthetic ($\varepsilon=5.94$, no filt.) &  $71.5\%$ ($\pm$0.40) \\
\hline
Synthetic ($\varepsilon=2.91$) & $73.2\%$ ($\pm$0.48) \\
\hline
Synthetic ($\varepsilon=5.98$) & $74.0\%$ ($\pm 0.62$) \\
\hline
Synthetic (300K, $\varepsilon=5.98$) & $\textbf{74.5\%}$ ($\pm 0.41$) \\
\hline
        \end{tabular}
    \end{adjustbox}
\end{table}

Table~\ref{tbl:winrate_arena} presents the win-rates for various models evaluated on 800 random instructions from the Chatbot Arena test split. The baseline model is LLaMA 7B fine-tuned with 180 thousand FLAN instructions. We also run evaluations using the baseline model trained on non-private Chatbot Arena data, with  the results presented in Appendix~\ref{apdx:sft_nonprivate}. The findings are consistent with those in Table~\ref{tbl:winrate_arena}.    We include  Vicuna-v1.3 \citep{vicuna2023} in our evaluation, one of the state-of-the-art  models that is non-privately fine-tuned from LLaMA on conversations collected from the ShartGPT website \citep{ShareGPT}.  The results in Table~\ref{tbl:winrate_arena} suggest two main findings. First, private synthetic instructions generated by our algorithms are of high utility, only showing a minor performance drop when compared to using real instructions without any privacy protection. Second, fine-tuning the models on the domain of private instructions yields significantly better performance compared to fine-tuning the models on out-of-domain FLAN instructions. In addition to  Chatbot Arena instructions, we also evaluate the models on the AlpacaEval benchmark \citep{dubois2023alpacafarm}, which  comprises 805 instructions collected from public evaluation sets. We present the results in Appendix~\ref{apdx:sft_aplcaeval}. It is noteworthy that  AlpacaEval instructions do not accurately represent real-world user queries. Nonetheless, such evaluation allows  us to make direct comparisons with  models listed on the AlpacaEval leaderboard \citep{alpaca_eval}. When evaluated on AlpacaEval, the best performing DP models slightly underperform Vicuna-v1.3 but are still comparable with other leading opensource  models such as LLaMA2-Chat \citep{touvron2023llama2}.

In Table~\ref{tbl:winrate_arena}, we also present the results of private training with real instructions.   As the real instructions are not privately released, the corresponding responses are also sensitive. Therefore, we use DP-Adam when training on pairs of real instructions and responses. In contrast, when training with synthetic instructions that have  been privately released, we can use standard optimizers because of the post-processing property of DP. It is important to note that even if we use DP-Adam for real instructions, instructions  containing personal information are not protected from annotators.  As indicated in Table~\ref{tbl:winrate_arena}, private training with real instructions yields inferior performance compared to  training with synthetic instructions. We hypothesize that this is because privatizing responses brings extra cost in model performance. In Figure~\ref{fig:len_instrut_vs_response} in Appendix~\ref{apdx:additional_plots}, we show responses are typically much longer than instructions, making privatizing them a more challenging task.

\subsection{RLHF with Proximal Policy Optimization}
\label{subsec:rlhf}

\begin{figure*}[htb]
\centering
\begin{subfigure}{.49\textwidth}
    \centering
    \includegraphics[width=1.0\linewidth]{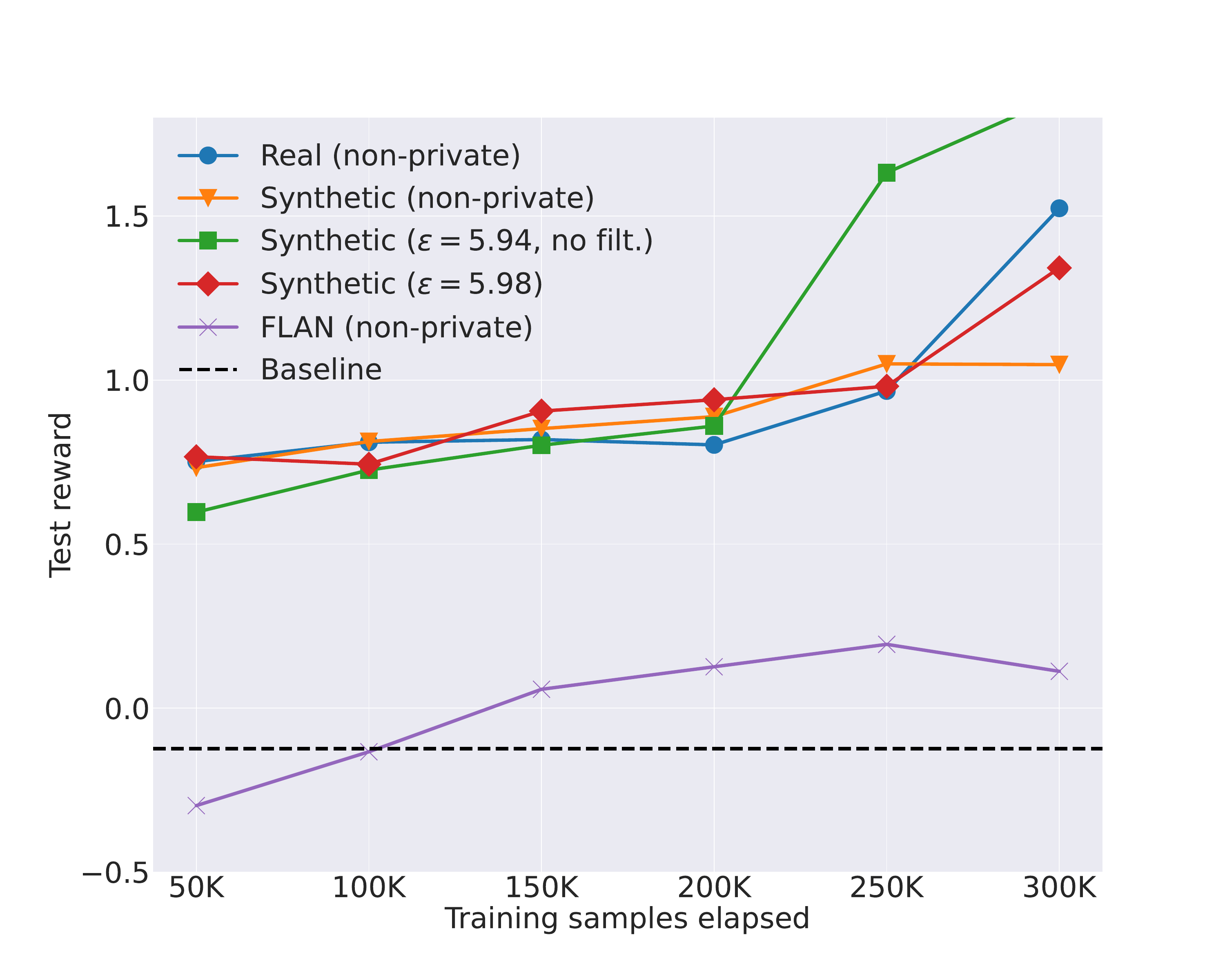}  
    \caption{Average test reward.}
\end{subfigure}
\begin{subfigure}{.49\textwidth}
    \centering
    \includegraphics[width=1.0\linewidth]{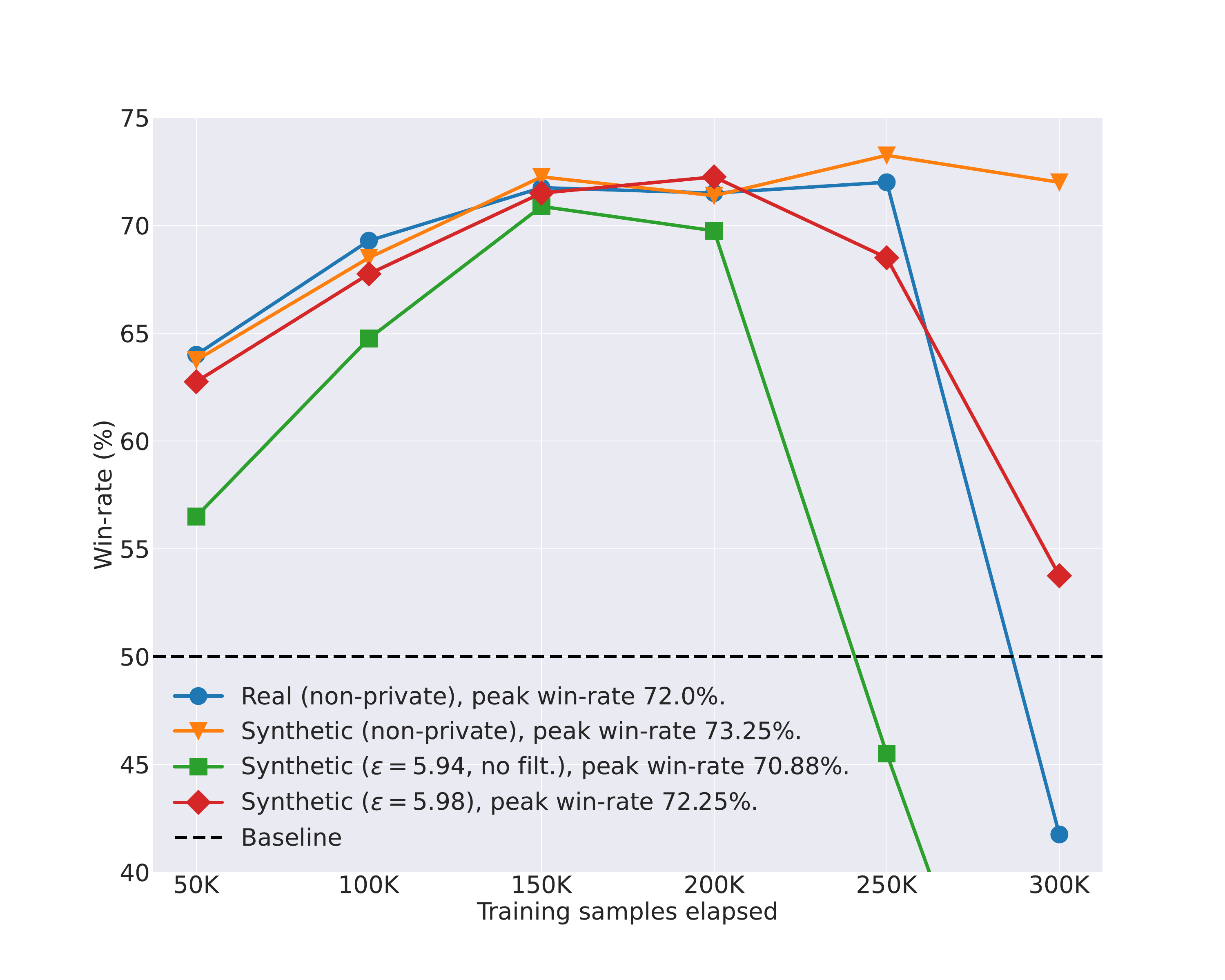}  
    \caption{Win-rate against the baseline model.}
\end{subfigure}
\caption{Model performance along the PPO training trajectory. Using synthetic instructions yields similar reward and win-rate curves as using real instructions. In contrast, using non-private but out-of-distribution FLAN instructions yields inferior performance. For real instructions and private synthetic instructions, there is a decline in instruction-following ability in the later stages due to reward over-optimization \citep{gao2023scaling}.  }
\label{fig:result_ppo}
\end{figure*}

In this section, we evaluate the utility of our synthetic instructions in Reinforcement Learning from Human Feedback (RLHF). RLHF starts with a baseline model generating initial responses to a set of unlabeled instructions. The responses are then evaluated using some reward models trained to simulate human preference. Then, the baseline model is optimized through RL to increase the reward signal. We use Proximal Policy Optimization (PPO) as the RL algorithm \citep{schulman2017proximal,ouyang2022training}. Ideally, the reward model should be developed through private training using (synthetic) real-world user instructions and human-assessed rewards. However, due to the cost of developing reward models, we utilize an off-the-shelf model that comprises 304M parameters and is trained on  public human preference datasets. The baseline model  is fine-tuned from phi-1.5 \citep{li2023textbooks}  on the OpenOrca dataset \citep{OpenOrca}. Phi-1.5 has 1.3B parameters, which is smaller than the sizes of LLaMA models. We opt for a smaller model because the policy rollout stage in RL is time-consuming for large models.  More details regarding the PPO implementation are  documented in Appendix~\ref{apdx:hyperparameters_ppo}.

Our RLHF experiments use four distinct types of data: (1) Chatbot Arena instructions, (2) Synthetic instructions generated by a non-private model, (3) Private synthetic instructions (with/without filtering), and (4) FLAN instructions. We run PPO for 300 thousand instructions. In the cases of synthetic instructions and FLAN instructions, we randomly sample 300 thousand instructions from the corresponding datasets. This means that no instruction is reused during PPO training. In the case of Chatbot Arena instructions, the training makes more than one pass on the training data since its training split contains only  180 thousand samples.

We evaluate checkpoints along the PPO training trajectory, using 800 random instructions from the Chatbot Arena test split. The results are presented in Figure~\ref{fig:result_ppo}. Our evaluation focuses on two key aspects: the average test rewards and win-rates against the baseline model. We use GPT-4-Turbo as the judge and employ the prompt from \citet{alpaca_eval} to compute win-rates. For both real and private synthetic instructions, we observe an increase in rewards but a decline in instruction-following ability in the later stages of training. This phenomenon is known as reward over-optimization because the reward model serves as an imperfect proxy for actual human preferences, which is an active research topic~\citep{gao2023scaling}. The results, as depicted in Figure~\ref{fig:result_ppo}, demonstrate that the reward and win-rate curves for our synthetic instructions closely mimic those achieved using real instructions. This underscores the high utility of our synthetic instructions.

\section{Conclusion}
\label{sec:conclusion}

For the common practice of fine-tuning LLMs on instructions, we identify a new threat model on the user provided instructions: potential risks in exposing sensitive personal information in the instructions to human annotators. To address this concern, as well as preventing fine-tuned LLMs to memorize sensitive information in user instructions, we propose to replace user instructions with DP synthetic instructions and introduce a novel framework for generating high-quality instructions by using a DP histogram to resample synthetic instructions to match the distribution of the real dataset in the embedding space. The effectiveness of our framework  is validated through comprehensive experiments on publicly available Chatbot Arena datasets~\citep{zheng2023lmsys,ChatbotArenaConversations} and LLaMA and Phi-1.5 models~\citep{touvron2023llama1, li2023textbooks}. Important future directions include developing algorithms capable of privately learning from user instructions in multiple modalities, including image and speech.

\section*{Acknowledgement}

The authors thank Krishna Pillutla for the insightful discussions on Algorithm~\ref{alg:dp_filt}, and Natalia Ponomareva for her valuable comments and constructive feedback on an early draft.

\section*{Impact Statement}
In Appendix~\ref{sec:instructions_are_sensitive}, we describe a method for identifying sensitive information within instructions. Our goal is to demonstrate the vulnerability of real-world user inputs, in the hope of encouraging leading AI companies to prioritize the protection of user privacy. While it is possible that potential adversaries might employ the described method, we argue that the negative societal impact of this work is minimal. The method we employ is standard, and the instructions under analysis are already public.

\bibliography{contents/ppml}
\bibliographystyle{contents/sty_files/bst_file}

\onecolumn

\appendix
\section{Additional Related Work}
\label{apdx:more_related_work}

\citet{wu2024privately} present pioneering efforts in aligning language models with differential privacy. They align GPT-2 models \citep{radford2019language} using private optimizers to ensure that models trained with both supervised fine-tuning and reinforcement learning satisfy differential privacy. They demonstrate that the rewards achieved by DP models are comparable to those of non-private models on the IMDb dataset \citep{maas2011learning} and the Reddit TL;DR dataset \citep{volske2017tl}. Compared to \citet{wu2024privately}, this work introduces several new aspects. First, our framework mitigates privacy risks arising from both human annotation and model memorization. In contrast, the privacy guarantees in \citet{wu2024privately} only apply to the trained models. Second, our experiments focus on instruction datasets collected during the real-world deployment of LLMs \citep{zheng2023lmsys}. Moreover, we evaluate the aligned models' ability to follow instructions not only through rewards but also through win-rates evaluated by LLM judges \citep{zheng2023judging,dubois2023alpacafarm}.

\citet{lin2023differentially} and \citet{xie2024differentially} investigate generating differentially private synthetic data using only API access to foundation models. 
Their approach, termed Private Evolution, also leverages votes from real samples to select synthetic samples.
Private Evolution is a multi-round generation algorithm. 
In each round, it first generates a pool of candidate synthetic samples by applying variation APIs to samples from the previous round. Then they sample from the candidate pool based on votes from real datapoints. The key difference between their selection process with ours is that we cluster synthetic samples to improve the signal-to-noise ratio during the selection process. Our Algorithm~\ref{alg:dp_filt} allows us to run the selection process with a minimal privacy cost. Additionally, we focus on training instruction-following models and conduct comprehensive experiments to verify the effectiveness of our algorithms.

Recently, \citet{zhang2023examining} show that users may disclose private information in their conversations with LLMs. They investigate the conversations collected on the ShareGPT website. In this work, we  investigate the user instructions in the  Chatbot Arena Conversations dataset\footnote{\url{https://huggingface.co/datasets/lmsys/chatbot_arena_conversations}.}. Compared to the analysis in \citet{zhang2023examining}, our analysis is new in two aspects. First, all conversations from ShareGPT are  proactively shared by the user themselves and hence may result in an inherent sample bias. In contrast, users instructions from Chatbot Arena are directly collected by the service provider. Second, we conduct a quantitative analysis of six distinct categories of personally identifiable information, providing a more comprehensive understanding for the presence of common types of sensitive information.

In addition to DP synthetic text, significant research has been dedicated to generating other forms of data with differential privacy, including synthetic tabular data \citep{jordon2018pate,tantipongpipat2019differentially,ge2020kamino,tao2021benchmarking,rosenblatt2020differentially,aydore2021differentially,vietri2020new,vietri2022private,mckenna2021winning,mckenna2022aim,liu2023generating} and synthetic image data \citep{xie2018differentially,torkzadehmahani2019dp,harder2021dp,dockhorn2022differentially,ghalebikesabi2023differentially,sehwag2023differentially}. In the realm of generating DP synthetic tabular data, recent studies have explored enhancing the quality of synthetic data through  post-processing the initial outputs \citep{neunhoeffer2020private,liu2021iterative,wang2023post}. Most recently, \citet{wang2023post} propose to privatize the correlation matrix of a private dataset, and then re-sample the initial synthetic data based on the privatized correlation matrix. Their approach achieves state-of-the-art performance on several benchmarks.  To the best of our knowledge, this work marks the first attempt to investigate the post-processing of private synthetic text. We introduce a filtering algorithm that demonstrates both high effectiveness and minimal additional privacy costs.

Leveraging pre-trained models in private learning has substantially improved the privacy-utility trade-off of deep learning with DP. This advancement extends to various domains, including natural language processing \citep{majmudar2022differentially,bu2023differentially,wang2023can,du2023dp,xiao2023large,duan2023flocks,tang2023privacy,tang2024private,panda2023differentially,hong2023dp}, computer vision \citep{luo2021scalable,golatkar2022mixed,mehta2022large,panda2022dp,de2022unlocking,zhu2023private,tang2023differentially,xu2023learning,thaker2023leveraging,zhang2023differentially,tobaben2023efficacy}, and speech recognition \citep{shamsabadi2022differentially,azam2023importance}. Recent theoretical studies have provided evidence to explain the efficacy of using pre-trained models in private learning \citep{li2022does, ganesh2023public}. In this paper, we follow this line of research and utilize DP fine-tuning to train generative models for creating synthetic instructions. It is noteworthy that if the pre-training process does not satisfy differential privacy, then DP fine-tuning only assures the privacy of fine-tuning data. In real-world deployments, it is important to also consider the privacy risks of pre-training data \citep{tramer2022considerations}. One potential solution is to also enforce differential privacy in the pre-training stage \citep{kurakin2022toward,anil2022large,sander2022tan}.

\newpage

\section{Real-world User Instructions Are Sensitive}
\label{sec:instructions_are_sensitive}

\begin{figure*}[ht]
\centering
\begin{subfigure}{.4\textwidth}
    \centering
    \includegraphics[width=1.0\linewidth]{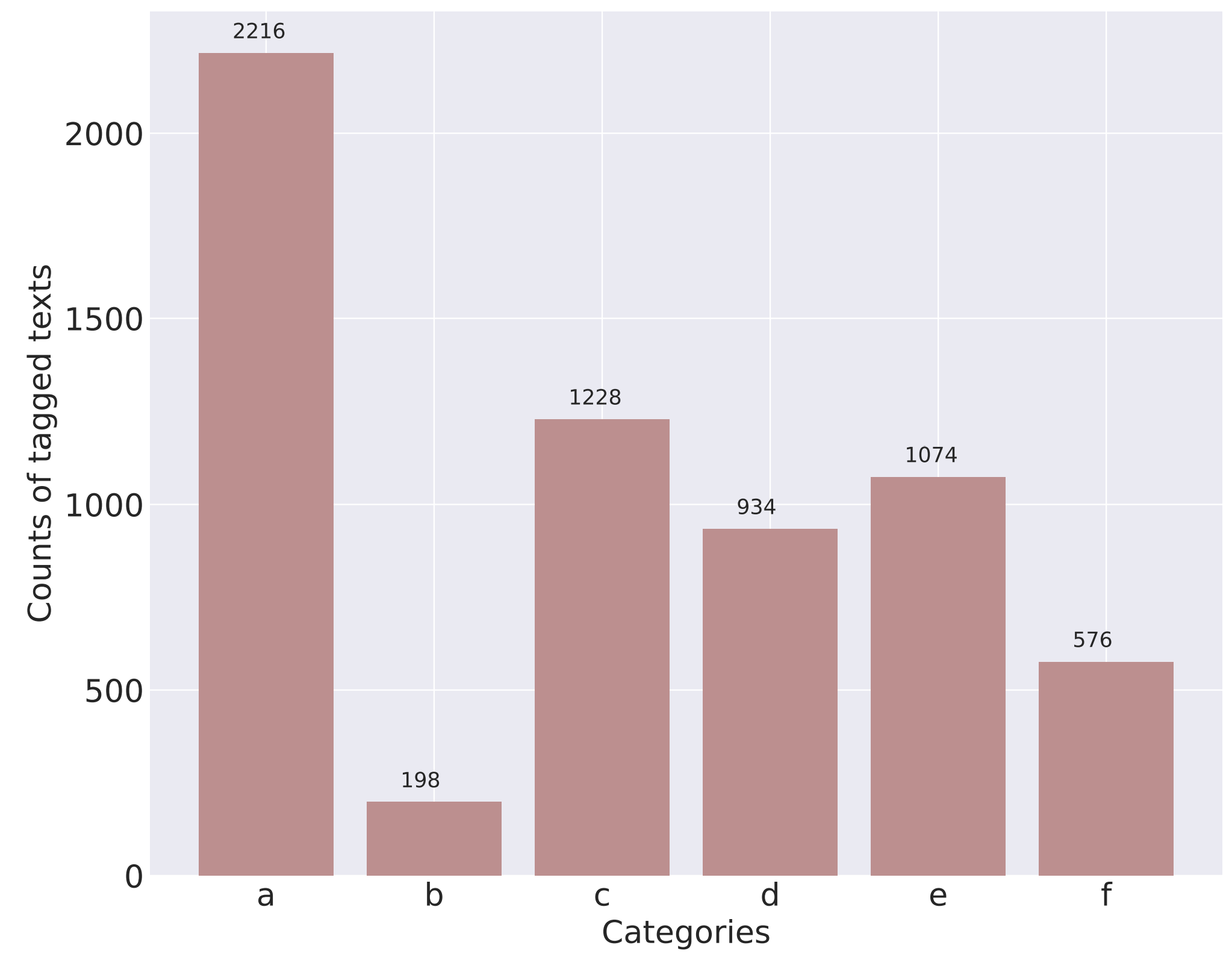}  
    \caption{Counts of PII by category in the initial screening.}
    \label{fig:initial_screen_category_count}
\end{subfigure}
\begin{subfigure}{.4\textwidth}
    \centering
    \includegraphics[width=1.0\linewidth]{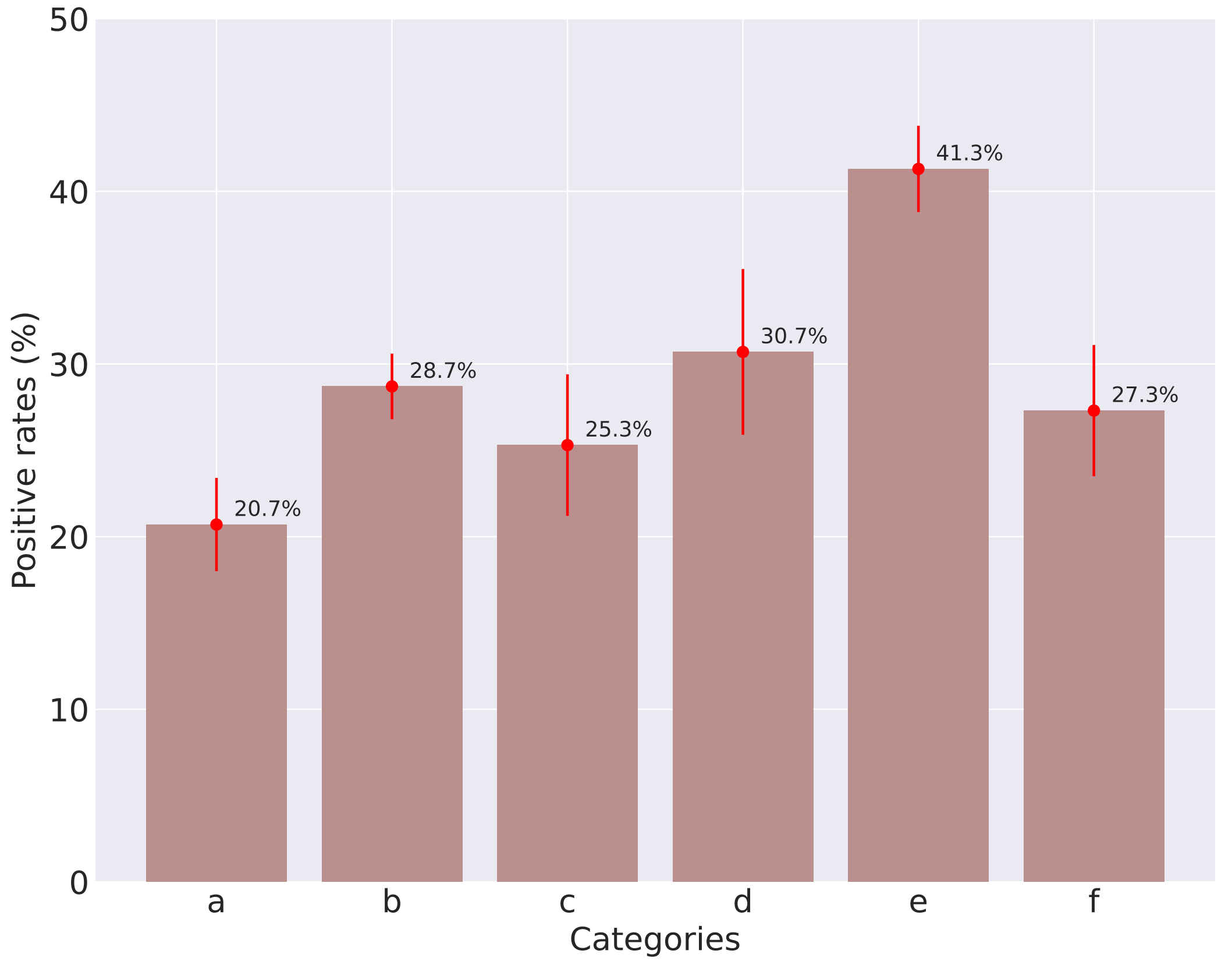}  
    \caption{True positive rates in human annotation.}
    \label{fig:results_human_labelling}
\end{subfigure}
\caption{Main results of our PII analysis. The x-axis represents six categories of sensitive text pieces: a. full names of real individuals, b. unique identifiers such as email addresses, c. specific locations, d. names of organisations, e. occupational titles, and f. specific times. Figure (a) shows the results of initial screening by prompting GPT-3.5-Turbo to flag sensitive text pieces in user instructions. Figure (b) presents the true positive rates of these initial screening results, as determined by human annotation.  }
\label{fig:pii_analysis}
\end{figure*}

We conduct a systematic analysis to demonstrate the prevalence of private information in real-world user instructions. We investigate around 12 thousand instructions from the Chatbot Arena Conversations dataset~\citep{ChatbotArenaConversations}. The dataset has 33 thousand conversations with human preferences collected on the Chatbot Arena website, and each conversation sample contains a user instruction, two responses from two different models, and preference from the user. In our analysis, we only use the user instructions, and concatenate the instructions of different rounds into one string for multi-round conversations. We deduplicate the instructions and only keep instructions that have more than 20 GPT-3.5-Turbo tokens, which reduces the number of instructions from 33 thousand to 12,472 instructions.

\begin{figure*}[htb]
\begin{AIbox}[width=1.0\textwidth]{}
\parbox{1\textwidth}{\scriptsize\begin{alltt} 

1. For the given instruction from a user (denoted by INSTRUCTION), identify the following personally identifiable information.

	a. The full names of real individuals.
	
	b. Number or code that identifies entities in real word, such as phone number, email address, personal or organization website.
	
	c. Specific places and locations, such as: cities, areas, named infrastructures (bus stops, bridges, etc.).
	
	d. Names of organisations, such as companies, schools, universities.
	
	e. Occupational titles, such as job names, position names.
	
	f. Description of a specific time, such as October 3, 2018 or 13th June.
	
	
2. List the answer in the following format: [[catergory]]: personally identifiable information.

3. If a personally identifiable information is mentioned multiple times, only count it once.

4. Do not include personally identifiable information in public articles (such as news) or fiction stories.



DEMONSTRATIONS:  ...


INSTRUCTION = ...

RESULT:

\end{alltt}}
\end{AIbox}
\caption{The prompt for initial screening with GPT-3.5-Turbo. The categories of sensitive texts are inspired by the categories of personally identifiable information in \citet{pilan2022text}. The authors label two emails from the Enron Email dataset as demonstrations. }
\label{fig:gpt35_prompt}
\end{figure*}

\begin{figure}[htb]
\centering
\includegraphics[width=0.5\textwidth]{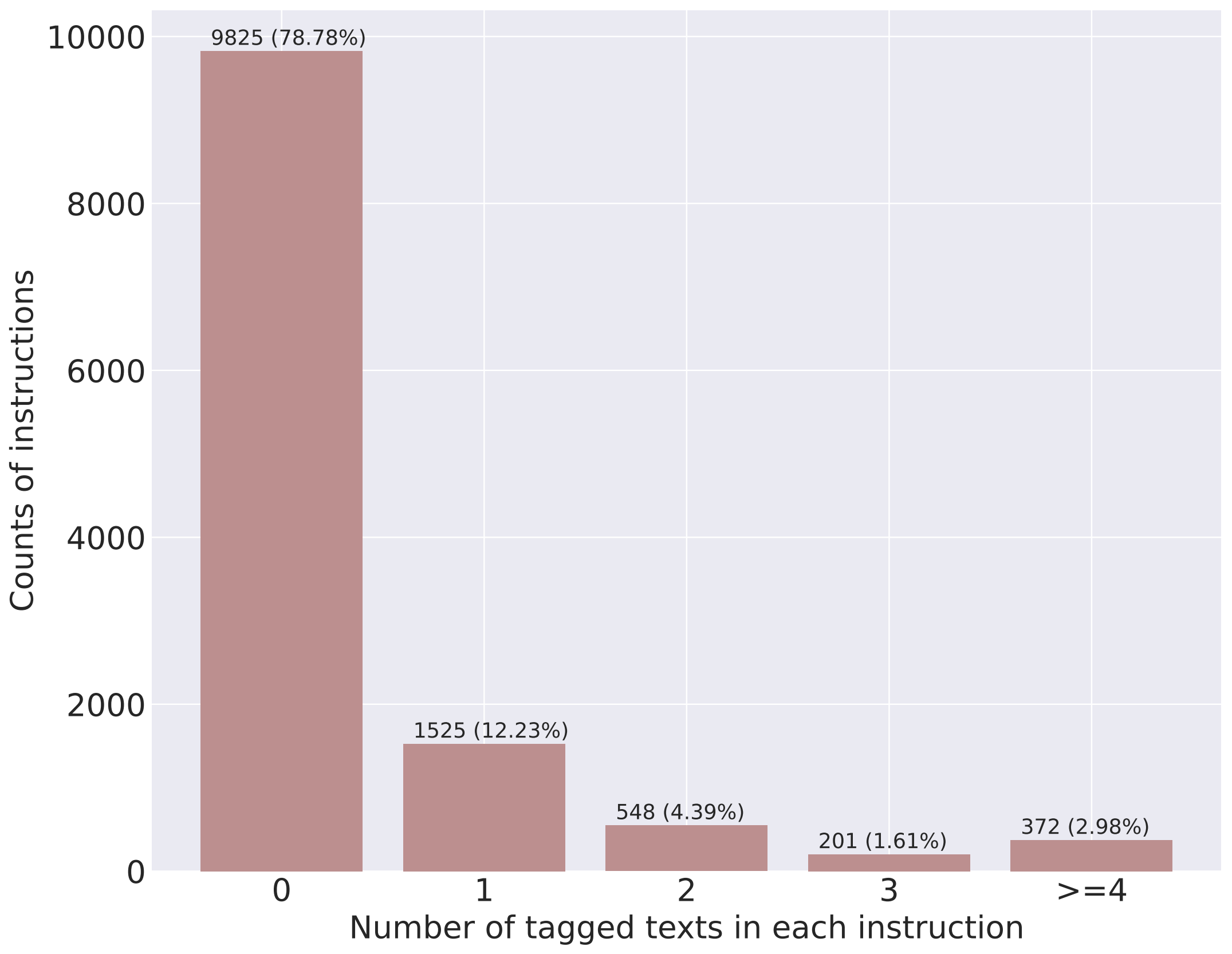}
\caption{The number of instructions containing varying numbers of text pieces that are flagged by GPT-3.5-Turbo as sensitive.
}
\label{fig:initial_screen_instruction_count}
\end{figure}

 Inspired by the personally identifiable information (PII) categories in \citet{pilan2022text}, we consider six categories of sensitive text pieces, including: a. full names of real individuals, b. unique identifiers such as email addresses, c. specific locations, d. names of organisations, e. occupational titles, and f. specific times. We first prompt GPT-3.5-Turbo to flag text pieces that might be sensitive as the first round of screening. We choose LLM-based PII detection instead of rule-based detection because recent studies suggest LLMs achieve a higher detection rate for PII than existing rule-based detection tools \citep{bubeck2023sparks}. We show the prompt used for LLM-based PII detection in Figure~\ref{fig:gpt35_prompt} and the  counts of flagged text pieces by category in Figure~\ref{fig:initial_screen_category_count}. Out of the 12,472 instructions being analyzed, more than 6000 text pieces in total, ranging in length from a single word to a few words, are flagged by GPT-3.5, and more than 20\% of the instructions contain at least one piece of sensitive text. The histogram of the number of instructions containing varying numbers of flagged text pieces is presented in Figure~\ref{fig:initial_screen_instruction_count}.

\begin{figure*}[htb]
\begin{AIbox}[width=1.0\textwidth]{}
\parbox{1\textwidth}{\scriptsize\begin{alltt} 


\textbf{Annotation guide}: Below is a piece of information tagged as sensitive by GPT-3.5, along with its accompanying instruction. Does it appear to relate to real individuals and should be considered as sensitive? Do NOT try to identify the user in the real world.


\textbf{Tagged text}: business CEOs

\textbf{Instruction}: are creative jobs, \hl{business CEOs}, and social jobs safe from being replaced by AGI? explain thoroughly.

\textbf{Label}: not sensitive.


\textbf{Tagged text}: Marketing Director

\textbf{Instruction}: Please give me a profile summary based on my work experience: currently \hl{[Job title and company name]} in \hl{[Location]}. Previously 19 years in \hl{[Company name]}, promoted from \hl{[Job title]} to \hl{[Job title]} and then to \hl{Marketing Director} for ...... 

\textbf{Label}: sensitive.

\end{alltt}}
\end{AIbox}
\caption{Annotation guideline and sample annotations made by the authors. Sensitive text pieces, expected for the one that is being annotated, are redacted by the authors to protect user privacy.}
\label{fig:example_human_labelling}
\end{figure*}

\begin{figure*}[htb]
\begin{AIbox}[width=1.0\textwidth]{}
\parbox{1\textwidth}{\scriptsize\begin{alltt} 


\textbf{Type}: ChatGPT makes false predictions.

\textbf{Sample instruction}: Which of the following is NOT a component of the 1995 \hl{Dodge Viper} mechanical system? A: Toaster oven, B: Flywheel, C: Clutch, D: Exhaust headers


\textbf{Type}: The entities are either public figures or fictional characters.

\textbf{Sample instruction}: Write a compare and contrast essay about \hl{Tom Buchanan} and \hl{Jay Gatsby} from \hl{F. Scott Fitzgerald's} \hl{The Great Gatsby}.



\textbf{Type}: The entities are from a public news.

\textbf{Sample instruction}: translate to spanish

"""
\hl{(Reuters)} - \hl{International Business Machines Corp} expects to pause hiring for roles as roughly 7,800 jobs ......, CEO \hl{Arvind Krishna} told \hl{Bloomberg News} on Monday.
"""

\end{alltt}}
\end{AIbox}
\caption{Common types of false positive sensitive pieces that are flagged by GPT-3.5 but then annotated as not sensitive by the authors.}
\label{fig:false_positives}
\end{figure*}

While the initial screening by LLM marks many text pieces as sensitive, we find that many of the flagged text pieces are false positives that are not concerning. For instance, names in fictional stories or public news are often identified as sensitive by GPT-3.5, even if we instruct the LLM to not flag them. We manually check the flagged text pieces to estimate the true positive rates. We randomly sample 50 flagged pieces from each of the six categories. And for each tagged text, the annotators label whether it is sensitive (true positive) or not. The annotation guideline and two sample annotations are shown in Figure~\ref{fig:example_human_labelling}. We repeat this process for three times with different random seeds for sampling and 150 flagged text pieces are manually reviewed for each category. The average positive rates and standard deviations are in Figure~\ref{fig:results_human_labelling}. The average positive rate is larger than 20\% for every category. The category with the lowest and highest positive rates are `full names of real individuals' (20.7\%) and `occupational titles' (41.3\%), respectively. In Figure~\ref{fig:false_positives}, we present common types of false positive predictions made by GPT-3.5. 
Examples of sensitive instructions found are shown in Figure~\ref{fig:true_positives}.

\newpage

\section{Empirical Privacy Leakage}
\label{sec:canary}

\begin{figure*}[htb]
\begin{AIbox}[width=1.0\textwidth]{}
\parbox{1\textwidth}{\scriptsize\begin{alltt} 
\textbf{Repetition}: 1

\textbf{Canary}: Make the ending of this email more friendly.

Thanks in advance for your help! Looking forward to hearing from you.

Thanks,

Jay

Tel: 716-802-3549

\textbf{Canary}: Convert this address into json format {'street':, 'city':,'state':}.

172 Homestead Street, Manchester, CT.


\textbf{Repetition}: 10

\textbf{Canary}: Which state does this phone number belong to?

Number: (253) 697-4180

\textbf{Canary}: Create a travel itinerary to Canada. Include one night stay at my friend's place: 531 Bobbin Mill Road, Lunenburg, VT.


\textbf{Repetition}: 100

\textbf{Canary}: Does this text message sound persuasive?

Please call us at 643 827 1509 ASAP otherwise you will lose your discount.

\textbf{Canary}: I didn't receive my package, what should I reply to this email?

Your package to 2715 Thornbrook Court, Odenton, MD has been delivered. Please contact us if you need assistance.
\end{alltt}}
\end{AIbox}
\caption{Six canary instructions. Each canary contains either a fake phone number or a fake address. }
\label{fig:canary_instructions}
\end{figure*}

One common empirical approach to assess the privacy risk of a training algorithm is injecting canaries  (specifically crafted training samples) into the training set, and evaluating the extent to which the canaries are exposed through querying the trained model \citep{pillutla2023unleashing,nasr2023tight,jagielski2023note}. In this section, we follow this approach to compare the empirical privacy leakage of models trained with and without DP. We use six canary instructions, three of them contain phone numbers and the other three contain addresses (see Figure~\ref{fig:canary_instructions} for the six canaries).  Phone numbers are randomized 10-digit sequences, and addresses are generated using an off-the-shelf tool\footnote{\url{https://pypi.org/project/random-address}.}. Following \citet{yue2022synthetic}, we repeat the canaries with different times (1, 10, and 100) to explore the worst case privacy leakage \citep{lee2021deduplicating,kandpal2022deduplicating}. It is important to note that the privacy parameter $\varepsilon$ for a canary linearly increases with the number of repetitions \citep{algofound, vadhan2017complexity}.

\begin{table}[htb]
\renewcommand{\arraystretch}{1.1}
\centering
    \caption{Evaluating the leakage of injected canaries. `Loss Rank' is the rank of a canary among 10,000 instructions that use the same template but different secrets. The rightmost two columns indicate if the trained model leaks the secrets during inference, conducted in two ways. First, we generate 100,000 synthetic instructions without prompts and check for the presence of secrets. Second, we use text preceding the secret as prompt to see if the model's greedy decoding leaks the secret.  } \label{tbl:empirical_leak}
      \smallskip
    \begin{adjustbox}{max width=0.75\textwidth}
        \begin{tabular}{c|c|c|c|c}
\hline
Instance-level $\varepsilon$  &  Repetition & Loss Rank & Leakage (No Prompt) & Leakage (With Prompt) \\ 
\hline
\multirow{3}{*}{Non-private} & 1    &   50     & 0/2      &  0/2    \\
                             & 10 &  1    & 1/2          &   2/2\\
                            & 100 &  1    & 2/2         &   2/2 \\
\hline
\multirow{3}{*}{$\varepsilon=2.86$} & 1    &   4142    & \multirow{3}{*}{0/2}     &  \multirow{3}{*}{0/2}  \\
                                    & 10 &  1232    &  &  \\
                                    & 100 & 806    &   &  \\
\hline
\multirow{3}{*}{$\varepsilon=5.94$} & 1    &   3566  &  \multirow{3}{*}{0/2}  & \multirow{3}{*}{0/2} \\
                & 10 &  1254    &   &   \\
                & 100 &  698    &   &  \\
\hline
        \end{tabular}
        
    \end{adjustbox}
   
\end{table}

We quantify the leakage  using two measures. The first measure is the relative position of a canary's loss among the losses of 10,000 instructions. These instructions follow the same template but differ in the phone numbers or addresses they contain. For each repetition count, we calculate two loss ranks: one for a canary with a phone number and another for a canary with an address. We only report the lower of these two ranks. The second measure evaluates whether the trained model  leaks the secret verbatim during inference. We run inference  both with and without prompts. For running inference without prompt, we generate 100,000 synthetic instructions using each trained model and check for the presence of secrets in the generations. The hyperparameters used for sampling are the same as those used in Section~\ref{sec:exps}. For running inference with prompts, we utilize the text preceding the secret as the prompt and examine if the model's completion, through greedy decoding, contains the secret.  Table~\ref{tbl:empirical_leak} presents the leakage of injected secrets. For models trained without DP, the loss ranks of all inserted canaries are alarmingly low, indicating that these models memorize the injected secrets. Furthermore, non-private models leak the secrets when canaries are repeated for 10 or 100 times, whereas private models demonstrate no such leakage.


\section{Ablation Studies}
\label{apdx:ablation_studies}

\subsection{The Impact of Gradient Clipping Thresholds on MAUVE Scores}
\label{apdx:vary_clipping}

\begin{figure}[h!]
\centering
\includegraphics[width=0.5\textwidth]{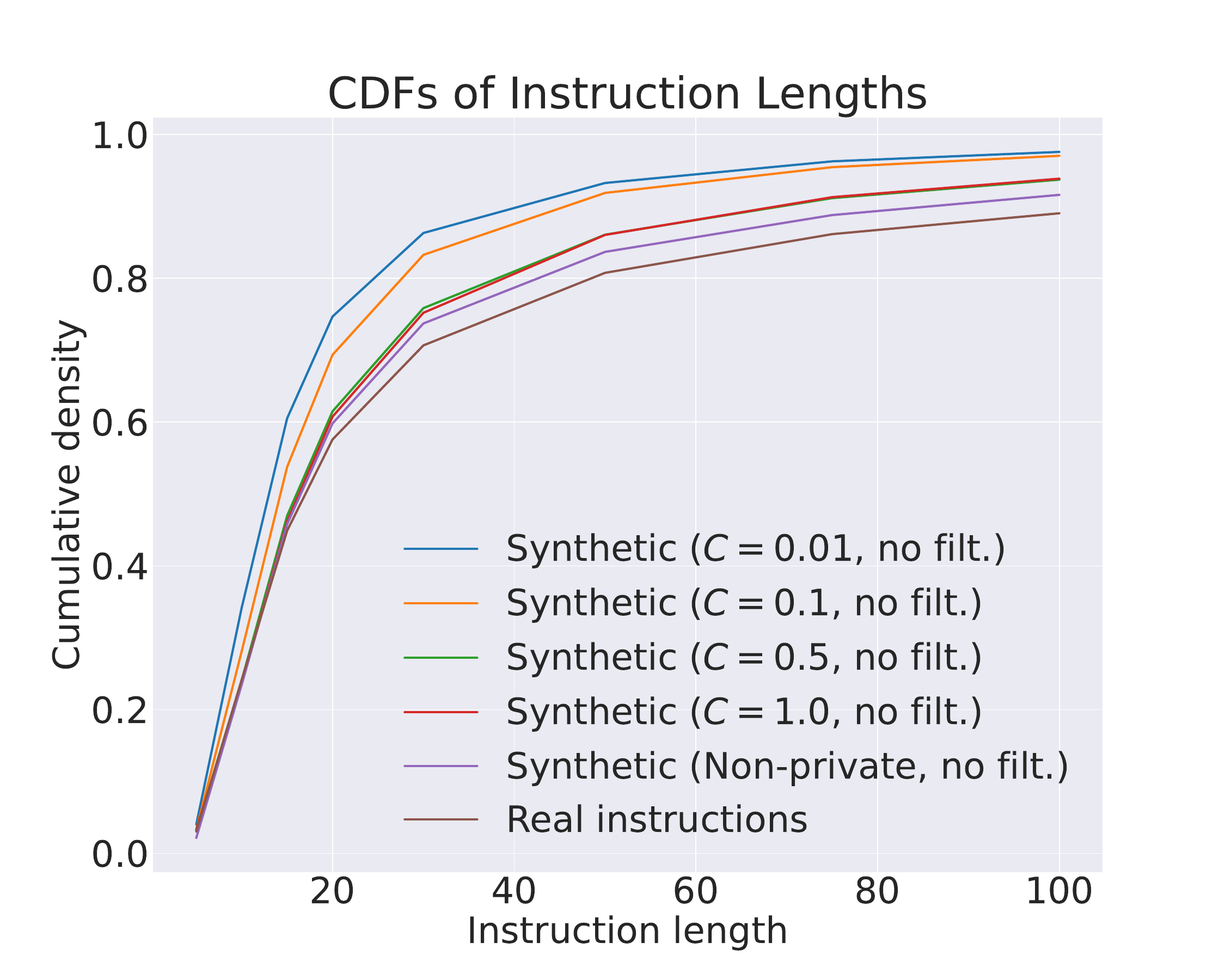}
\caption{CDFs of instruction length distributions. The generators are trained with different gradient clipping thresholds $C$.  The synthetic instructions are sampled from 7B generative models and are not filtered with Algorithm~\ref{alg:dp_filt}. The privacy budget is $(5.94,5\times 10^{-7})$-DP. Small values of $C$ bias the trained models to generate short instructions.
}
\label{fig:vary_clipping}
\end{figure}

 \begin{table}[h!]

\renewcommand{\arraystretch}{1.1}
\centering
    \caption{The influence of gradient clipping thresholds on MAUVE scores. Models are fine-tuned from LLaMA-7B with $\varepsilon=5.94$. The instructions are not filtered with Algorithm~\ref{alg:dp_filt}.  } \label{tbl:mauve_clip}
      \smallskip
    \begin{adjustbox}{max width=0.45\textwidth}
        \begin{tabular}{l|c|c}
\hline

    & Unigram & Sentence-T5 \\
\hline
 Non-private  &    $0.980$     &      $0.987$         \\
\hline
$C=0.01$  &     $0.667$    & $0.843$ \\
\hline 
$C=0.1$ &   $0.704$      &  $0.874$ \\
\hline 
$C=0.5$  &  $\textbf{0.839}$       & $\textbf{0.890}$ \\
\hline
$C=1.0$  &   $0.819$      & $0.868$ \\
\hline
        \end{tabular}
    \end{adjustbox}
\end{table}

We find that the gradient clipping threshold in DP fine-tuning, denoted as $C$, influences the quality of the initial synthetic instructions. A smaller value of $C$ makes the fine-tuned generative models bias toward producing shorter instructions. This phenomenon likely occurs because the loss for a sequence is the sum of  all token losses in that sequence\footnote{We normalize the sequence loss by the maximum sequence length to prevent loss explosion. Another common practice is to take the average of token losses as the sequence loss. However, our preliminary experiments suggest that taking the average loss biases short sequences regardless the choice of clipping threshold.}. As a result, longer sequences tend to have higher losses and larger gradients, causing a smaller $C$ to bias the model towards shorter sequences. Conversely, a larger $C$ introduces higher noise variance, which can also degrade the quality of the fine-tuned models. Figure~\ref{fig:vary_clipping} displays the cumulative distribution functions of  instruction lengths for varying $C$ values. Table~\ref{tbl:mauve_clip} presents the MAUVE scores. We set $C=0.5$ in the rest of this paper as we find it strikes a balance between these two effects.

\subsection{Larger Pre-trained Models Achieve Better MAUVE Scores}
\label{apdx:llama7b_vs_13b}

Table~\ref{tbl:mauve_7b_vs_13b} presents a comparison between using LLaMA 7B and LLaMA 13B as the pre-trained models. We find that using LLaMA 13B as the pre-train model leads to better MAUVE scores. This aligns with the notion that larger pre-trained models improves DP fine-tuning, as evidenced by recent studies \citep{ganesh2023public,he2023exploring,berrada2023unlocking}.

\begin{table}[htb]
\renewcommand{\arraystretch}{1.1}

\centering
    \caption{MAUVE scores of synthetic datasets generated by 7B and 13B models. The synthetic data is not filtered with Algorithm~\ref{alg:dp_filt}. Fine-tuning 13B models gives better synthetic data. } \label{tbl:mauve_7b_vs_13b}
      \smallskip
    \begin{adjustbox}{max width=0.45\textwidth}
        \begin{tabular}{l|c|c}
\hline

    & Uni-gram & Sentence-T5 \\
\hline
 Non-private (7B)  &     $0.980$    &   $0.987$            \\
\hline
 Non-private (13B)  &     $0.983$    & $0.991$ \\
\hline 
$\varepsilon=5.94$ (7B) &    $0.839$     & $0.890$ \\
\hline 
$\varepsilon=2.86$ (13B) &  $0.932$       & $0.893$ \\
\hline
$\varepsilon=5.94$ (13B) &  $\textbf{0.942}$        & $\textbf{0.912}$ \\
\hline
        \end{tabular}
    \end{adjustbox}
\end{table}

\subsection{Run Algorithm~\ref{alg:dp_filt} with Fewer Initial Synthetic Instructions}
\label{apdx:fewer_initial}

\begin{figure*}[htb]
\centering
\begin{subfigure}{.425\textwidth}
    \centering
    \includegraphics[width=1.0\linewidth]{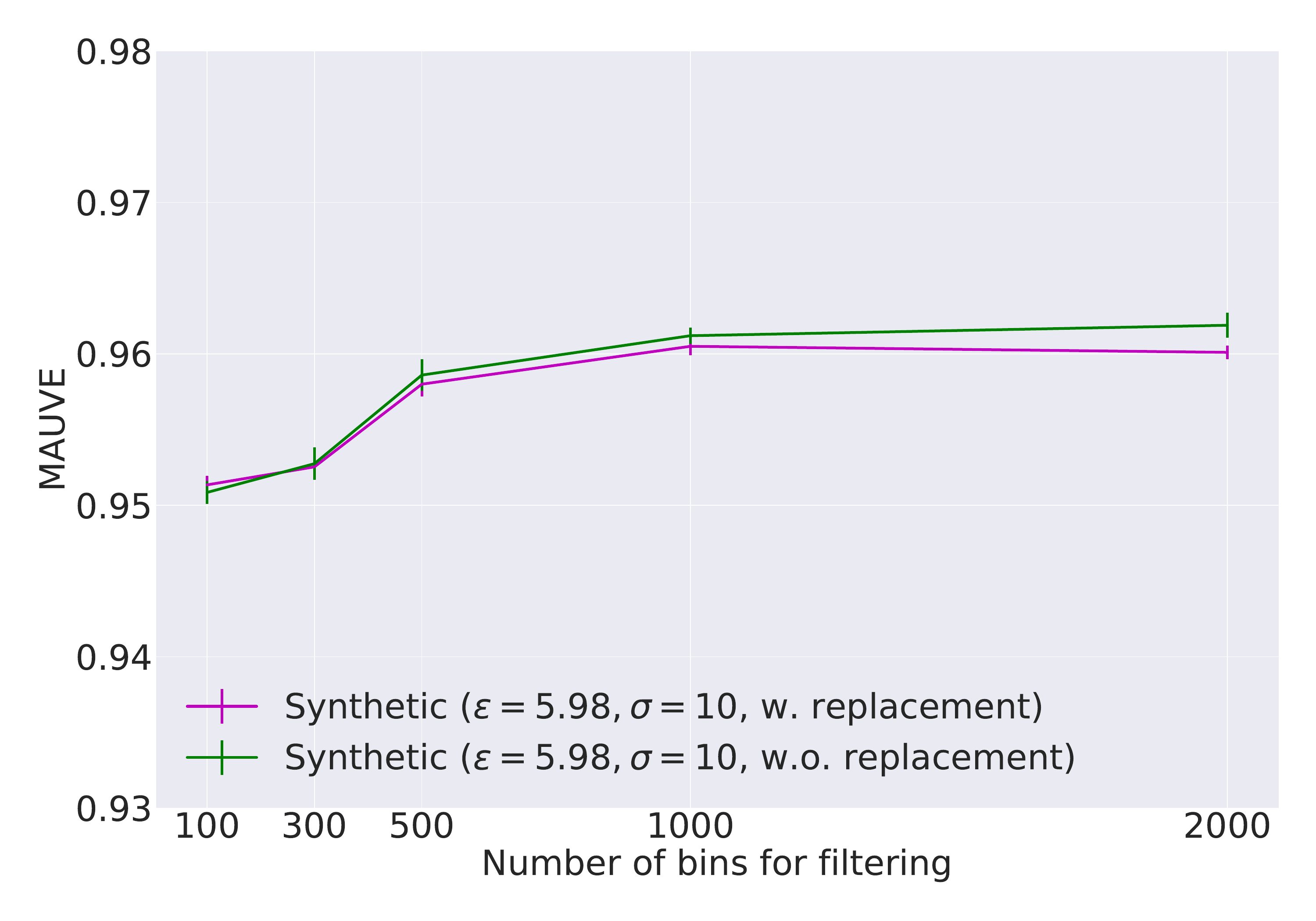}  
    \caption{MAUVE-Unigram}
\end{subfigure}
\begin{subfigure}{.425\textwidth}
    \centering
    \includegraphics[width=1.0\linewidth]{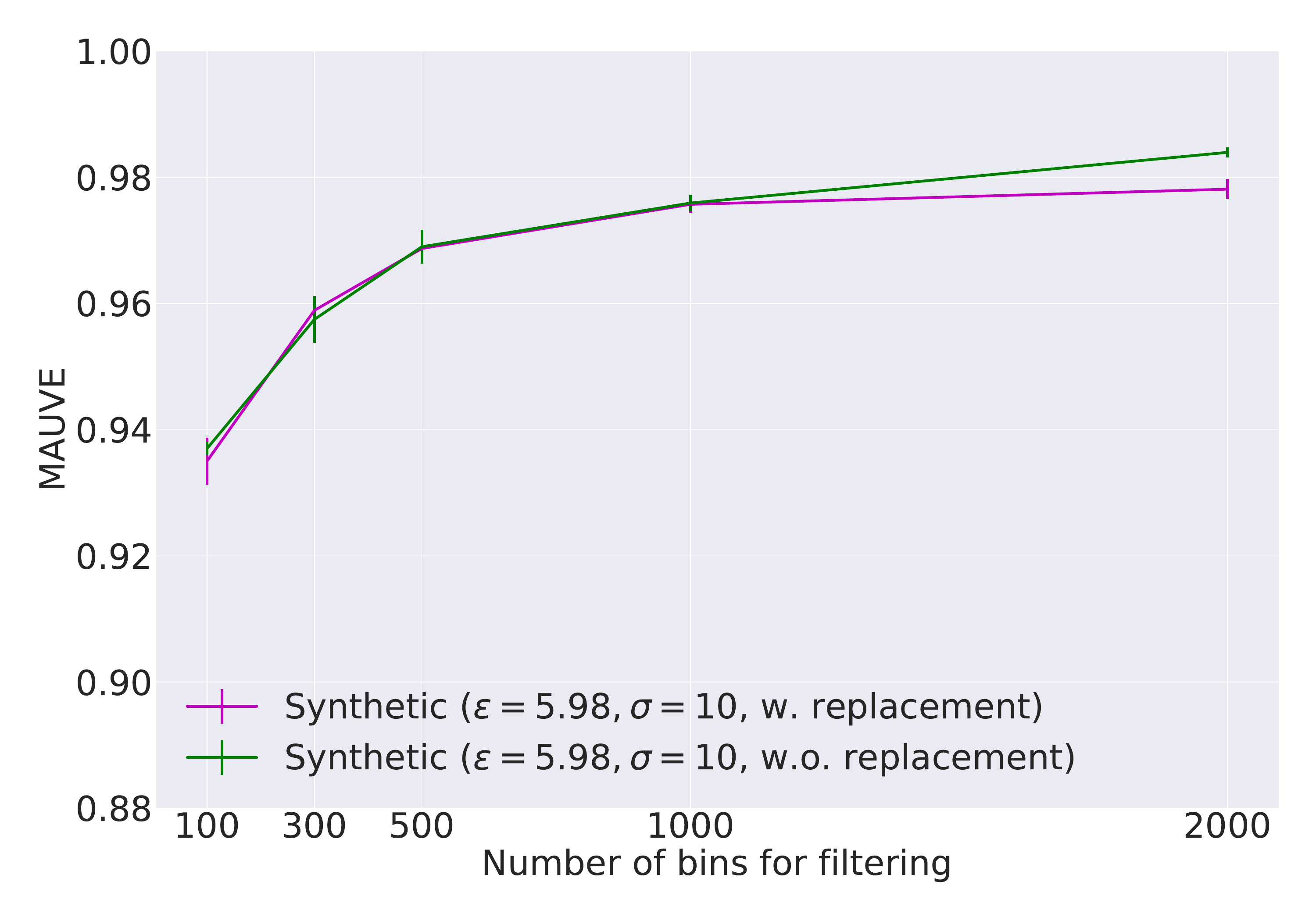}  
    \caption{MAUVE-Sentence-T5}
\end{subfigure}
\caption{Re-sampling from 500 thousand or one million  initial synthetic instructions. For filtering 500 thousand instructions, sampling with replacement is used for $K\geq 1000$ otherwise there are not enough initial samples to select a subset of 180 thousand. }
\label{fig:sample_w_replacement}
\end{figure*}

In our experiments in Section~\ref{sec:exps}, we run Algorithm~\ref{alg:dp_filt} to filter one million synthetic instructions. To investigate the effect of using fewer initial synthetic instructions, we present the MAUVE scores of filtering 500 thousand initial synthetic instructions in Figure~\ref{fig:sample_w_replacement}. It is noteworthy that for $K=1000$ and $K=2000$, we use sampling with replacement during the re-sampling process in Algorithm~\ref{alg:dp_filt} because otherwise there are not enough initial samples to select from. The results in Figure~\ref{fig:sample_w_replacement} indicate that for large values of $K$, using fewer initial synthetic instructions results in worse MAUVE scores.

\subsection{Run Algorithm~\ref{alg:dp_filt} on More Datasets}
\label{apdx:other_datasets}

We use two more datasets to show the wide applicability of Algorithm~\ref{alg:dp_filt}. The datasets are 75,000 IMDB movie reviews\footnote{\url{https://huggingface.co/datasets/stanfordnlp/imdb}.} and 75,316 PubMed paper abstracts\footnote{\url{https://huggingface.co/datasets/pubmed}.} (from August 01 to August 05, 2023). To reduce computational cost, the generators are fine-tuned from GPT2-Large. The hyperparameters for DP fine-tuning follow the best ones in \citet{yue2022synthetic}. When Algorithm~\ref{alg:dp_filt} is applied, we increase the noise multiplier for DP fine-tuning to compensate for the privacy cost of the re-sampling process. We generate 750,000 initial synthetic samples and select a subset that is of the same size as real data. Algorithm~\ref{alg:dp_filt} consistently improves the quality of initial synthetic data, as indicated by the MAUVE scores in Table~\ref{tbl:more_datasets}.

\begin{table}[htb]
\renewcommand{\arraystretch}{1.1}

\centering
    \caption{MAUVE scores (Sentence-T5) of synthetic IMDB reviews and PubMed abstracts. Generators are fine-tuned from GPT2-Large. } \label{tbl:more_datasets}
      \smallskip
    \begin{adjustbox}{max width=0.45\textwidth}
        \begin{tabular}{l|c|c}
\hline

    & IMDB & PubMed \\
\hline
Synthetic ($\varepsilon=3.96$, no filitering)  &     $0.647$    &   $0.594$            \\
\hline
Synthetic ($\varepsilon=3.93$)  &     $0.825$    & $0.811$ \\
\hline 
        \end{tabular}
    \end{adjustbox}
\end{table}

\subsection{Evaluate SFT Models Using the Non-private Baseline}
\label{apdx:sft_nonprivate}

We evaluate the win rates using a baseline model trained on non-private Chatbot Arena data. The results are presented in Table~\ref{tbl:winrate_non-private}. These findings align with those presented in Table~\ref{tbl:winrate_arena} of Section~\ref{subsec:sft}, where the baseline model is trained on FLAN data. The results show that: 1) the proposed filtering algorithm improves the win-rates; 2) the best-performing DP model are comparable to the non-private baseline. 

\begin{table}[htb]
\renewcommand{\arraystretch}{1.1}
\centering
    \caption{Win-rates with the baseline model trained on non-private Chatbot Arena data.} \label{tbl:winrate_non-private}
      \smallskip
    \begin{adjustbox}{max width=0.45\textwidth}
        \begin{tabular}{l|c}
\hline

\multicolumn{2}{c}{\textbf{7B Models}} \\
\hline
 Chatbot Arena (non-private) & $50\%$   \\ 
\hline
Vicuna-v1.3  &  $46.0\%$ ($\pm$0.20)    \\
\hline 
Synthetic ($\varepsilon= 5.94$, no filt.) &  $44.2\%$ ($\pm$0.27)  \\
\hline 
Synthetic ($\varepsilon= 5.98$) &  $48.0\%$ ($\pm$0.62)  \\
\hline 
 Synthetic (300K, $\varepsilon= 5.98$) &  $49.1\%$ ($\pm$0.51)   \\
\hline
\hline
\multicolumn{2}{c}{\textbf{13B Models}} \\
\hline
Vicuna-v1.3 &  $57.7\%$ ($\pm$0.66)   \\
\hline
Synthetic (300K, $\varepsilon=5.98$) &  $59.2\%$ ($\pm$0.41)  \\
\hline
        \end{tabular}
    \end{adjustbox}
\end{table}

\subsection{Evaluate SFT Models on AlpacaEval}
\label{apdx:sft_aplcaeval}

\begin{table}[h!]
\renewcommand{\arraystretch}{1.1}
\centering
    \caption{Win-rates evaluated on AlpacaEval. The best performance of DP models are marked in bold. The best DP models are comparable with LLaMA2-Chat but slightly underperform Vicuna-v1.3.} \label{tbl:winrate_alpaca}
      \smallskip
    \begin{adjustbox}{max width=0.45\textwidth}
        \begin{tabular}{l|c}
\hline

\multicolumn{2}{c}{\textbf{7B Models}} \\
\hline
Text-davinci-003  & $50\%$  \\ 
\hline
FLAN (non-private)  &  $63.9\%$    \\
\hline
LLaMA2-Chat  &  $71.4\%$ \\ 
\hline
Vicuna-v1.3  &  $76.8\%$    \\
\hline 
Chatbot Arena (non-private) &   $74.2\%$ \\
\hline 
Chatbot Arena ($\varepsilon=5.94$) &  $64.9\%$  \\
\hline 
Synthetic ($\varepsilon=5.94$, no filt.) &  $70.2\%$  \\
\hline 
Synthetic ($\varepsilon=5.98$) &  $\textbf{71.6\%}$ \\
\hline 
Synthetic (300K, $\varepsilon=5.98$) & $71.2\%$ \\
\hline
\hline
\multicolumn{2}{c}{\textbf{13B Models}} \\
\hline
LLaMA2-Chat  & $81.1\%$  \\ 
\hline
Vicuna-v1.3 &   $82.1\%$   \\
\hline
Synthetic (300K, $\varepsilon=5.98$) &  $\textbf{80.6\%}$  \\
\hline
        \end{tabular}
        
    \end{adjustbox}
   
\end{table}

In addition to the test set of Chatbot Arena instructions, we also evaluate the models in Section~\ref{subsec:sft} on the AlpacaEval dataset \citep{dubois2023alpacafarm}. AlpacaEval comprises 805 instructions collected from public evaluation sets, including MT-Bench \citep{zheng2023judging}, the self-instruct evaluation set \citep{wang2022self}, the Koala evaluation set \citep{geng2023koala}, etc. To prevent data contamination, we again apply an 8-gram deduplication between AlpacaEval instructions and Chatbot Arena instructions. We note that AlpacaEval instructions do not accurately represent real-world user queries to LLMs. For example, the self-instruct evaluation set is generated via prompting GPT-3. Nonetheless, we include the results for a more comprehensive evaluation. We use the same evaluation setup as that used by the AlpacaEval leaderboard\footnote{\url{https://tatsu-lab.github.io/alpaca_eval}.}, which allows  us to make direct comparisons with state-of-the-art models listed on the leaderboard. The results are presented in Table~\ref{tbl:winrate_alpaca}. When evaluated on AlpacaEval instructions, the best performing DP models slightly underperform Vicuna-v1.3 but are comparable with LLaMA2-Chat.

\section{Data Cleansing for the LMSYS-Chat-1M Dataset}
\label{apdx:data_preprocess}

This section details the data pre-processing procedures applied to the LMSYS-Chat-1M dataset \citep{zheng2023lmsys}. For the analysis in Appendix~\ref{sec:instructions_are_sensitive}, we utilize a separate dataset: the Chatbot Arena Conversations dataset\footnote{\url{https://huggingface.co/datasets/lmsys/chatbot_arena_conversations}.} \citep{zheng2023judging}. The pre-processing steps outlined below are not applied to it.

Specifically, for the LMSYS-Chat-1M dataset, we implement the following sequential pre-processing steps:

\textbf{Take single-round instructions from English conversations.} We first extract user instructions from the dataset, while discarding the responses from LLMs. For the purpose of simplifying our experimental setup, we only consider the instruction from the first round in case of multi-round conversations. \citet{zheng2023lmsys} provide a language tag for each conversation. We only take instructions from English conversations as English accounts for around 80\% of the instructions among over more than twenty languages.  Approximately 780,000 instructions reamain after this step. 

\textbf{Remove inappropriate instructions.} For each conversation in LMSYS-Chat-1M, \citet{zheng2023lmsys} utilize OpenAI's moderation API to provide a tag that indicates if a conversation contains inappropriate content. Additionally, they redact real person names in the conversations. We exclude conversations that are either flagged by the moderation API or contain redacted names. Approximately 500,000 instructions remain after this step.

\textbf{Sequence de-duplication.} Users might repeatedly ask the same question to elicit responses from different LLMs. If an instructions appears multiple times, we only keep it once. There are around 310,000 instructions left after this step.

\textbf{N-gram de-duplication.} We divide the instructions into training, validation, and test sets using random splits. To prevent data contamination, we implement an 10-gram de-duplication process. During this process, if an 10-gram is present in multiple instructions, then only the first instruction that has it will be kept. As a result, the dataset is reduced to approximately 232 thousand instructions. Additionally, since we use the Alpaca Eval instructions \citep{dubois2023alpacafarm} as one of our evaluation sets, we further exclude any instructions containing 10-grams that appear in the Alpaca Eval instructions. This leads to the removal of fewer than twenty instructions. The dataset is reduced to approximately 220,000 instructions after this step.

\textbf{Remove unusual repetitions.} Upon manually inspecting the remaining instructions, we find unusual repetitions of certain specific patterns. For instance, instructions following these two patterns occur over 16 thousand times.

\centerline{\small\textsl{`Write an instruction of [...] with [...] words in chemical industry'}}

\centerline{\small\textsl{`Write an article about the [...] [...] words in chemical industry'}}

After removing those unusual repetitions, there are around 200,000 instructions left. We take 180,000 as the training set, 5,000 as the validation set, and the rest as the test set.

\section{Implementation Details}
\label{apdx:implementation_details}

This section documents the implementation details of our experiments. 

\subsection{Setup for Computing MAUVE}
\label{apdx:compute_mauve}

We use the official implementation\footnote{\url{https://github.com/krishnap25/mauve}.} to compute MAUVE scores. For computing MAUVE with Sentence-T5 embeddings, we follow the default setup to quantize the embeddings into 500 bins. The absolute MAUVE scores are affected by a scaling constant $c$. Using a larger $c$ would result in a lower score. For computing MAUVE with unigram distributions, we use the default setup $c=5$. For computing MAUVE with Sentence-T5 embeddings, we set $c=10$ to better distinguish the scores of different datasets. We note that values of $c$ do not affect the ranking of different datasets.

\subsection{Hyperparameters for (DP) Fine-tuning}
\label{apdx:hyperparameters}

We employ parameter-efficient fine-tuning as it has been shown to reduce computational cost \citep{houlsby2019parameter} and improve the performance in private learning \citep{yu2022differentially, bu2022differentially, kurakin2023harnessing}. Specifically, we use LoRA fine-tuning \citep{hu2022lora}. Following the setup in previous studies \citep{hu2022lora, li2022large}, we use a  reparametrization rank of 8 and only apply LoRA to attention weights.  For 7B models, this results in 8.4M trainable parameters. For 13B models, this results in 13.1M trainable parameters.  We use the Adam optimizer \citep{kingma2015adam} with ($\beta_1$=0.9, $\beta_2$=0.999). Our experiments use 32 Nvidia A100 40G GPUs. We will open source our code.

\begin{table*}[ht]
\footnotesize
\centering 
\caption{Hyperparameters for DP fine-tuning. The text in bold indicates the hyperparameters that we eventually use.  } \label{tbl:hyperparameter_dp_finetuning}
\begin{tabular}{l c}
\toprule
    LR Schedule & Constant LR \\
    Weight decay & [\textbf{0}, 1e-4]  \\
    Batch Size & [1024, 2048, \textbf{4096}] \\
    Epochs & [3, \textbf{10}] \\
    Learning Rate & $[0.5, \textbf{1}, 3]\times 10^{-3}$ \\
    Clipping Threshold & [0.01, 0.1, \textbf{0.5}, 1.0] \\
\bottomrule
\end{tabular}
\end{table*}

\begin{table*}[ht]
\footnotesize
\centering 
\caption{Hyperparameters for non-private fine-tuning. The text in bold indicates the hyperparameters that we eventually use.  } \label{tbl:hyperparameter_non_private_finetuning}
\begin{tabular}{l c}
\toprule
    LR schedule & Constant LR \\
    Batch size & [32, \textbf{64}, 128] \\
    Epochs & [\textbf{3}, 10] \\
    Learning rate & $[1, \textbf{3}, 5, 10]\times 10^{-4}$ \\
\bottomrule
\end{tabular}
\end{table*}

Our hyperparameter sweeps for DP fine-tuning and non-private fine-tuning are outlined in Table~\ref{tbl:hyperparameter_dp_finetuning} and Table~\ref{tbl:hyperparameter_non_private_finetuning}, respectively. The hyperparameter tuning is done with LLaMA 7B.
The criteria for the final hyperparameters is achieving the best unigram MAUVE score. Due to computational constraints, we do not perform a full hyperparameter sweep. It is important to note that hyperparameter tuning in DP training does incur an additional privacy cost, albeit modest \citep{liu2019private, papernot2022hyperparameter, mohapatra2022role, ding2022revisiting,redberg2023improving}. To potentially mitigate this overhead, we replicate our hyperparameter tuning process using FLAN instructions as the private dataset. We find that the optimal hyperparameters mirror those of using Chatbot Arena instructions, with the exception of the clipping threshold (1.0 is the optimal value for FLAN). This suggests that tuning hyperparameters on public instructions and transfering the results to private instructions is one viable option for reducing the privacy cost associated with hyperparameter tuning.

\subsection{Hyperparameters for Sampling}
\label{apdx:hyperparameters_sampling}

We employ unconditional sampling to generate synthetic instructions from the fine-tuned models. Specifcially, we use Nucleus Sampling \citep{holtzman2019curious}, adhering to its default configuration where the probability cutoff parameter, top\textunderscore$p$ is set as 0.95. Following previous work \citet{yue2022synthetic,kurakin2023harnessing}, the sampling temperature $T$ is set as 1.0. In our preliminary experiments, we set $T$=0.7 and observe a significant deterioration in the MAUVE scores. The maximum sequence length for the generated instructions is 1024 tokens.

For supervised fine-tuning, we follow the prompt for Vicuna. The prompt template is \textsl{`A chat between a curious user and an artificial intelligence assistant. The assistant gives helpful, detailed, and polite answers to the user's questions. USER: [Instruction] ASSISTANT:'.} The maximum sequence length for the generated responses is set as 2048 tokens. During inference, we follow the default setup of Vicuna\footnote{\url{https://github.com/lm-sys/FastChat/tree/main}.}. Specifically, sampling temperature $T$ and probability cutoff parameter top\textunderscore$p$ are set as 0.7 and 1.0, respectively.

\subsection{Code Snippet for Privacy Accounting}
\label{apdx:code_snippet_privacy_accounting}

\begin{figure*}
\centering
\includegraphics[width=0.8\textwidth]{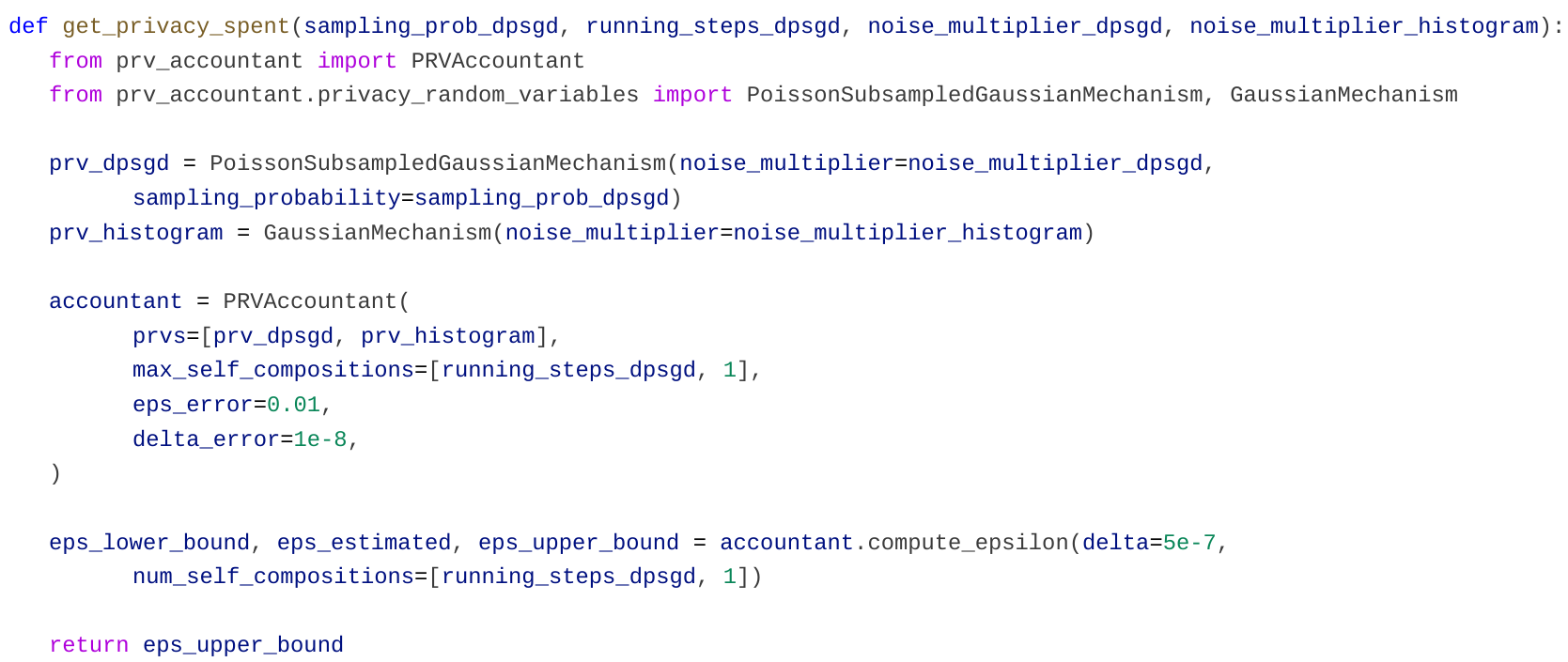}
\caption{The implementation of our privacy accounting function. We use the PRV accounting package provided by \citet{gopi2021numerical}.}
\label{fig:privacy_accounting_code}
\end{figure*}

The Python code of our privacy accounting function is in Figure~\ref{fig:privacy_accounting_code}. We use the Privacy Random Variable (PRV) accountant \citep{gopi2021numerical}. For running DP-Adam for 10 epochs with a batchsize of 4096 and $(5.94, 5\times 10^{-7})$-DP, we set the noise multiplier for DP-Adam as 0.81. For $(2.86, 5\times 10^{-7})$-DP, we set the noise multiplier as 1.11. The default noise multiplier for releasing the histogram in Algorithm~\ref{alg:dp_filt} is 10.0.

\subsection{Details for Annotating with GPT-3.5-Turbo}
\label{apdx:gpt_35_annotation}

We use gpt-3.5-turbo-0301\footnote{\url{https://platform.openai.com/docs/models/gpt-3-5}.} to create responses for all instructions. We use 20 parallel threads to query the OpenAI API. Annotating 100 thousand instructions takes approximately 12 hours and \$80. The maximum response length is 2048 tokens. The sampling temperature $T$ is set as 0.7. The probability cutoff parameter top\textunderscore$p$ is set to 0.95. 

\subsection{Hyperparameters for Proximal Policy Optimization}
\label{apdx:hyperparameters_ppo}

The baseline model\footnote{\url{https://huggingface.co/Open-Orca/oo-phi-1_5}.} is a 1.3B model fine-tuned on the OpenOrca dataset. The reward model\footnote{\url{https://huggingface.co/OpenAssistant/reward-model-deberta-v3-large-v2}.} comprises 304M parameters and is trained on several public human preference datasets, such as the annotated version of the TL;DR dataset featuring Reddit posts \citep{volske2017tl,stiennon2020learning}. The maximum sequence length for generated responses is set as 256 tokens. The sampling temperature and the probability cutoff parameter top\textunderscore$p$ are both set to 1.0 during training. Our PPO implementation is based on the Transformer Reinforcement Learning (TRL) library. We follow the default setup for PPO training\footnote{\url{https://github.com/huggingface/trl/blob/main/trl/trainer/ppo_config.py}.} for most of the hyperparameters. The KL penalty coefficient is set as 0.2 and the number of optimisation steps per batch is 4. We use a batchsize of 128 and a learning rate of $1.41\times 10^{-5}$. All models are trained for 2500 batches of samples. 

\section{Additional Figures}
\label{apdx:additional_plots}

\begin{figure}[htb]
\centering
\includegraphics[width=0.45\textwidth]{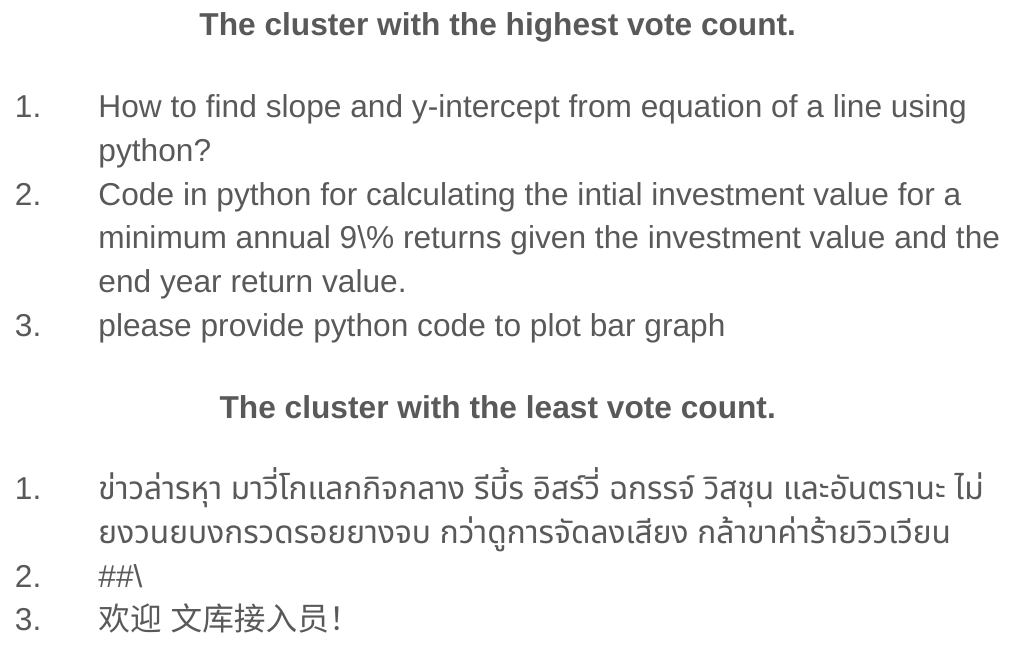}
\caption{Random synthetic instructions from clusters that have the highest and lowest vote counts.}
\label{fig:cluster_samples}
\end{figure}

\begin{figure}[htb]
\centering
\includegraphics[width=0.5\textwidth]{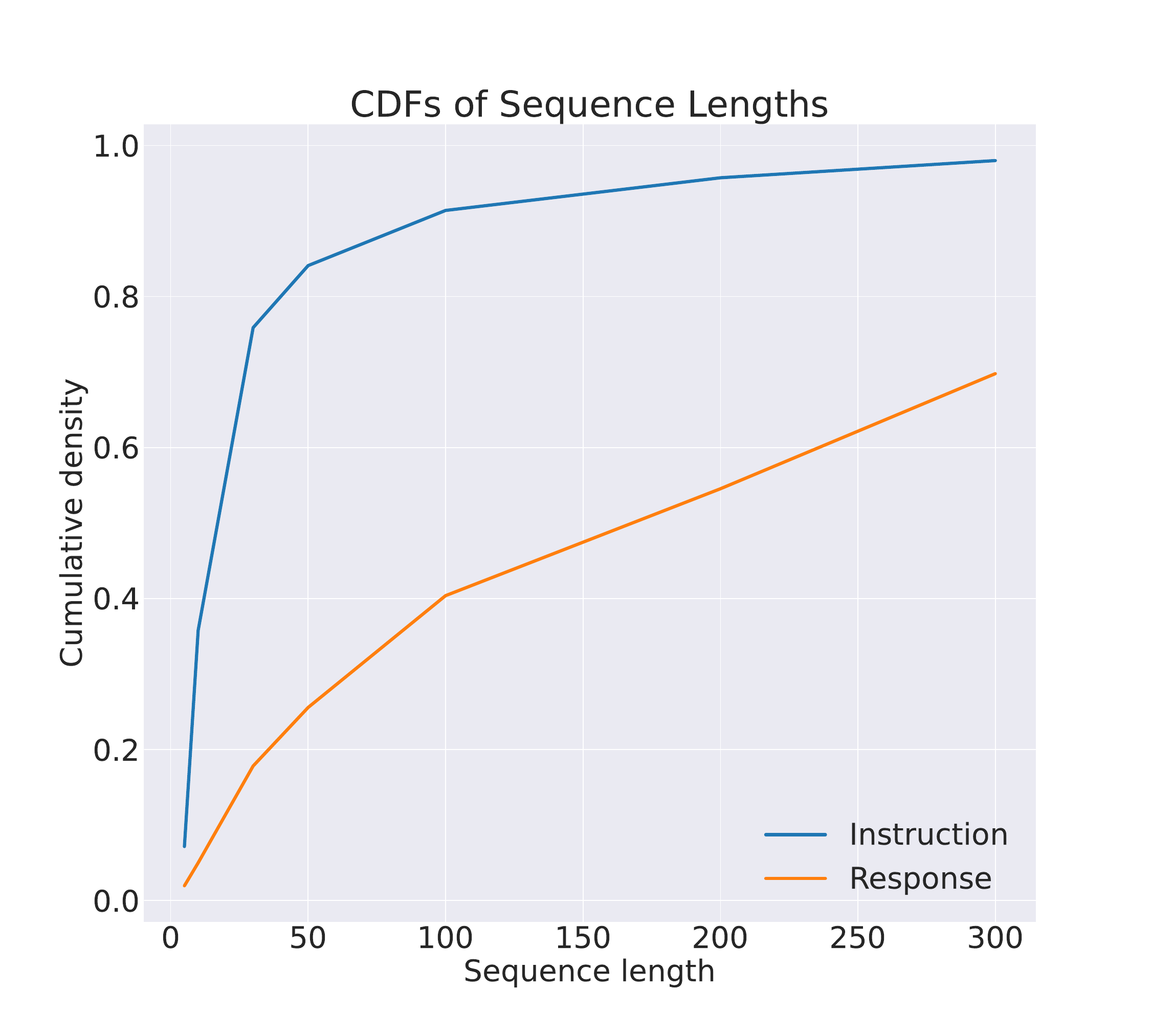}
\caption{Cumulative distribution functions of sequence length distributions. The responses are generated by GPT-3.5-Turbo. Responses are typically much longer than instructions.
}
\label{fig:len_instrut_vs_response}
\end{figure}

We put some figures here because of the space constraints.  In Figure~\ref{fig:cluster_samples}, we present random samples from two particular clusters of initial synthetic instructions: one cluster receiving the highest number of votes from real data and the other receiving the lowest. In Figure~\ref{fig:len_instrut_vs_response}, we compare the sequence length distribution of Chatbot Arena instructions with the length distribution of responses from GPT-3.5-Turbo.

\end{document}